\documentclass[a4paper,fleqn]{cas-sc}

\usepackage[numbers,sort&compress]{natbib}

\usepackage{graphicx,subfigure}

\newproof{proof}{Proof}
\newdefinition{remark}{Remark}

\def\tsc#1{\csdef{#1}{\textsc{\lowercase{#1}}\xspace}}
\tsc{WGM}
\tsc{QE}
\tsc{EP}
\tsc{PMS}
\tsc{BEC}
\tsc{DE}

\begin{document}
\let\WriteBookmarks\relax
\def\floatpagepagefraction{1}
\def\textpagefraction{.001}
\shorttitle{SOH prediction of LiBs based on SSA-BiGRU}
\shortauthors{J. Wen}

\title [mode = title]{State of health prediction of lithium-ion batteries for driving conditions based on full parameter domain sparrow search algorithm and dual-module bidirectional gated recurrent unit}                      
\tnotemark[1]

\tnotetext[1]{This work was supported by the Natural Science Foundation of Shanxi Province under Grant 202403021211088.}

\author[1]{Jie Wen}[
                        auid=000,bioid=1,
                        orcid=0000-0003-0302-4123]
\cormark[1]
\ead{wenjie015@gmail.com}

\credit{Conceptualization, Methodology, Investigation, Writing---original draft, Writing - Review \& Editing, Supervision, Project administration, Funding acquisition}

\affiliation[1]{organization={School of Electrical and Control Engineering, North University of China},
                city={Taiyuan},
                postcode={030051}, 
                state={},
                country={China}}

\author[1]{Chenyu Jia}
\credit{Conceptualization, Methodology, Investigation, Data Curation, Writing---original draft}

\author[1]{Guangshu Xia}
\credit{Software, Validation, Formal analysis, Resources}

%

%
%

\cortext[cor1]{Corresponding author}


\begin{abstract}
Aiming at the state of health (SOH) prediction of lithium-ion batteries (LiBs) for electric vehicles (EVs), this paper proposes a fusion model of a dual-module bidirectional gated recurrent unit (BiGRU) and sparrow search algorithm (SSA) with full parameter domain optimization. With the help of Spearman correlation analysis and ablation experiments, the indirect health indicator (HI) that can characterize the battery degradation is extracted first based on the incremental capacity (IC)  curves of the Oxford battery dataset, which simulates the driving conditions. On this basis, the filtered one-dimensional HI is inputted into the dual-module BiGRU for learning the pre- and post-textual information of the input sequence and extracting the sequence features. In order to combine the different hyperparameters in the dual-module BiGRU, SSA is used to optimize the hyperparameters in the full parameter domain. The proposed SSA-BiGRU model combines the advantages and structures of SSA and BiGRU to achieve the highly accurate SOH prediction of LiBs. Studies based on the Oxford battery dataset have shown that the SSA-BiGRU model has higher accuracy, better robustness and generalization ability. Moreover, the proposed  SSA-BiGRU model is tested on a real road-driven EV charging dataset and accurate SOH prediction are obtained.
\end{abstract}

%


\maketitle

\section{Introduction}
Since the 21st century, with the insufficient supply of traditional fossil fuel energy and the aggravation of environmental degradation, the demand for new energy has been increasing, and the scale of the new energy field has also risen \cite{bao2023global}. As one of the main energy storage devices with the advantages of high energy density, small size, and low cost, LiBs have profoundly influenced the development of transportation tools such as EVs as the main power source \cite{hu2021research,xia2023historical} and have also been widely used in electronic mobile devices, satellites, and grid energy storage systems \cite{wang2020comprehensive,dunn2011electrical}, etc. However, due to the characteristics of LiBs themselves, they will inevitably deteriorate or even be damaged during use, which can lead to the failure of the battery system and, in more serious cases, cause property damage and personal safety \cite{ren2019comparative,barre2013review,che2023health}. Therefore, accurate SOH prediction of LiBs is crucial in the prognostics and health management  of LiBs and battery management systems \cite{huang2023state}.

Along with the main electrochemical reactions, LiBs undergo complex and coupled internal side reactions during storage or charge/discharge, which causes degradation of the internal materials, such that LiBs degrades. In order to improve the stability and safety of battery systems in operation, there is an urgent need for high-precision SOH prediction of LiBs. Currently, studies on SOH prediction of LiBs are broadly divided into three categories: model-based methods, data-driven methods, and fusion methods \cite{hu2020battery,shu2021state}. 
Model-based methods are generally used for SOH prediction of LiBs by studying the degradation state inside the battery, building physical models, and using parameter estimation algorithms, which can be broadly categorized into three types: electrochemical models, equivalent circuit models and empirical models. In general, electrochemical models have the highest estimation accuracy and also the highest complexity. For example, Li et al. developed a low-order electrochemical model containing a single particle of the solid electrolyte interphase membrane formation set of active substances that swells and cleaves to enable fast capacity prediction for real-time SOH prediction \cite{li2018single}, while Hosseininasab et al. proposed a co-estimation scheme for fractional-order battery models based on the derivation of partial differential equations of the pseudo two dimensional model, which simultaneously achieves adaptive estimation of the capacity and internal resistance of the battery \cite{hosseininasab2022state}.
Although the model-based methods can reflect the degradation behavior and chemical properties of the battery under different operating conditions, the establishment of the prediction model requires an in-depth understanding of the internal structure or reaction mechanism of LiBs, which requires a large amount of a prior knowledge as a support, and relatively has a high degree of complexity, which is difficult to realize for non-field professionals.

Unlike model-based methods, data-driven methods based primarily on machine learning are more concerned with mining useful information from the historical degraded data of batteries without relying on a priori knowledge. In addition, the time spent on repetitive modeling can be greatly reduced due to the good portability of data-driven methods. Data-driven methods can be categorized into two types: direct prediction and indirect prediction. In direct prediction, the capacity and internal resistance of batteries are used as the prediction inputs, but they are difficult to obtain directly and the measurement instruments are very expensive. Indirect prediction refers to the use of voltage, current, temperature and other data of battery operation as features to train the prediction model, which leads to the SOH or capacity prediction of LiBs. Classical data-driven methods include relevance vector machine (RVM) \cite{liu2015health}, support vector machine (SVM) \cite{lyu2022synchronous,patil2015novel}, support vector regression (SVR) \cite{lin2022lithium,li2022state,wei2017remaining,chen2023capacity}, and artificial neural networks (ANN) \cite{zhang2019synchronous,tang2022indirect,tang2023health}, among which convolutional neural networks (CNN) has also been widely used in the field of SOH prediction for LiBs due to its good performance. For example, Yang hybridized 2D-CNN and 3D-CNN to estimate the relationship between features and cycles by fusing the charging voltage, current and temperature of the battery with a feature-focused algorithm and a multi-scale cycle-focused algorithm \cite{yang2021machine}. Meanwhile, recurrent neural networks (RNN) have been used to predict battery degradation due to the expertise in finding relationships between time series \cite{eddahech2012behavior}, and some variants of RNN have been widely used in this field. For instance, Luo et al. proposed a bidirectional long short-term memory (BiLSTM) network combined with a multi-scale hierarchical attention mechanism based on a hybrid data preprocessing method for features, and validated it on a battery dataset from a spacecraft, which effectively improves the prediction performance of multidimensional time series \cite{luo2023hybrid}. Fusion methods, on the other hand, use both model-based methods and data-driven methods to gain a combined advantage. For example, Obregon et al. proposed a convolutional autoencoder (CAE) combined with deep neural network for SOH online estimation framework, where CAE is used to extract features from electrochemical impedance spectroscopy in an unsupervised manner in \cite{obregon2023convolutional}.

According to the literature review, the SOH prediction of LiBs based on data-driven methods is getting more and more attention and recognition from both academia and industry due to the booming development of artificial intelligence, and we will be devoted to this aspect as well in this paper. Specifically, the SSA-BiGRU model will be designed for predicting SOH of LiBs under driving conditions. Currently, most of the data-driven methods favor feature extraction using the data as HI, such as voltage, current and temperature, which is still essentially a temporal feature expression and cannot provide in-depth analysis of the internal changes of LiBs based on the features themselves. Although there have been many studies using IC analysis and differential thermal capacity (DTC) analysis, more often than not, only a single piece of information is used as the HI, while other information is ignored. In this paper, the indirect HI that can characterize the battery degradation is extracted from the IC curve, and through Spearman correlation analysis and ablation experiments, the one-dimensional HI will be filtered and inputted into the dual-module BiGRU to extract the time series features.
On the other hand, most data-driven methods tend to follow empiricism in parameter settings when using machine learning or deep learning, which may lead to the construction of a prediction model that is not in an optimal state and lacks generalization and portability, resulting in it performing well only on a single dataset. To address this issue, SSA is used to perform hyperparameter optimization of the dual-module BiGRU in the full parameter domain in this paper, which allows the model to adaptively adjust the parameter structure based on the actual training data to enhance the flexibility and portability.
Moreover, most of the work on SOH prediction of LiBs based on data-driven methods has been performed on battery datasets generated under ideal experimental environments, such as the NASA battery dataset, the CALCE battery dataset, and so on. There are some differences between these datasets and the battery degradation data generated from actual working conditions, which makes the validated prediction models based on these datasets somewhat questionable and unknown in terms of applicability and credibility. To avoid this problem, the Oxford battery dataset that simulates driving conditions and is closer to real engineering applications is chosen, and the robustness of the proposed SSA-BiGRU model is checked on the charging data of real EVs in this paper. Therefore, the main contributions of this paper are as follows: (1) The indirect HI characterizing battery degradation is extracted based on the IC curve and screened with the help of Spearman correlation analysis and ablation experiments. (2) The dual-module BiGRU model for predicting SOH of LiBs is designed, and a full parameter domain optimization algorithm is adopted to optimize the parameters involved in the model performance separately and independently, breaking the symmetric parameter constraints of the traditional BiGRU. (3) The prediction performances of the SSA-BiGRU model are investigated using the dataset of simulated driving conditions and the charging dataset of real EVs.

The rest of this paper is organized as follows. The SOH, HI extraction method based on ICA, the used battery datasets and the evaluation indicators are described in Section \ref{sec:pre}. In Section \ref{sec:SSA-DBiGRU}, the proposed SSA-BiGRU model and the corresponding prediction process for SOH prediction of LiBs are presented. The application experiments based on two public battery datasets for the SSA-BiGRU model are performed and analyzed in Section \ref{sec:experment}. Finally, we conclude this paper and discuss the possible future works in Section \ref{Sec:Cons}.

\section{Preliminaries}\label{sec:pre}
\subsection{SOH}\label{subsec:soh}
Aging of LiBs is usually manifested by the degradation of usable capacity, and the capacity of LiBs is used to define the SOH in this paper, i.e., 
\begin{equation}\label{eq:soh}
	{\rm{SOH}}=\frac{C_{\rm{Reality}}}{C_{\rm{Nominal}}}
\end{equation}
where, $C_{\rm{Reality}}$ and $C_{\rm{Nominal}}$ represent the maximum available capacity and initial capacity of LiBs running up to the present time, respectively. SOH is mathematically represented as a scalar with an initial value of 1 and no units.
\subsection{HI extraction based on ICA}\label{subsec:ICA_HI}
For SOH prediction, accelerated capacity degradation is well handled by mapping battery capacity using HI. The IC curve describes the capacity change of the battery under the unit voltage during the charging and discharging process. Since the IC curve is discrete, it is difficult to extract its features directly, so it is necessary to filter the extracted IC curve (e.g. Savitzky-Golay filter) to obtain a smoother curve. 

\begin{figure}[!htbp]
	\centering
	\includegraphics[width=.5\textwidth]{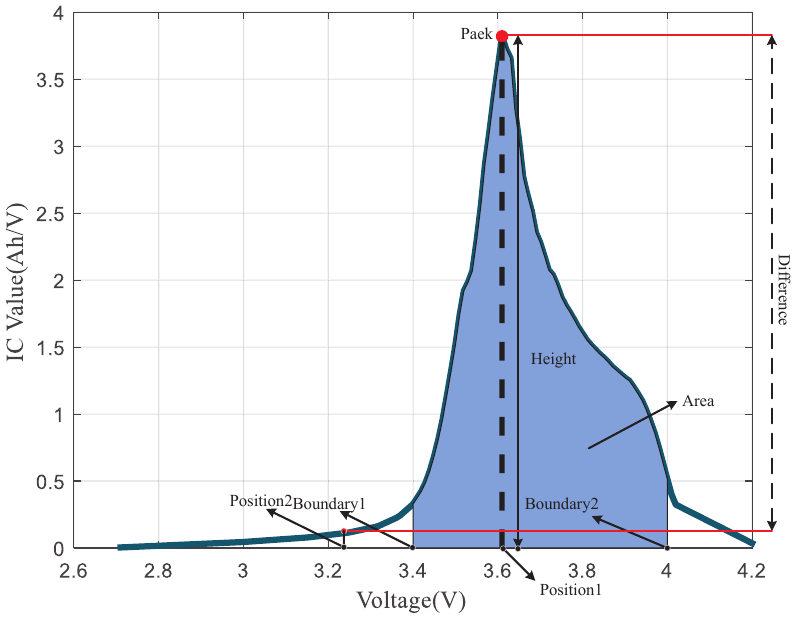}
	\caption{Schematic diagram of feature extraction based on IC curve.}
	\label{figur:ICA}
\end{figure}

In previous studies, researchers preferred to use the area under the IC curve or its peak value as the HI, and lacked other forms of information extraction. Different from it, in this paper, we first determines the voltage at which the IC curve reaches its peak, and then integrates the area in different voltage ranges and takes the highest correlation as the HI. The feature selection scheme is shown in Fig. \ref{figur:ICA}. Specifically, the peak of the IC curve is first determined to obtain the corresponding voltage point Position1($V$), and then the area under the curve is integrated according to the defined voltage intervals Boundary1($V$) and Boundary2($V$) , which is represented by the blue area in Fig. \ref{figur:ICA}. Moreover, the dimensionless features were tried to be extracted from each IC curve. In time domain analysis, the commonly used dimensionless features are crest factor (CF), pulse factor (PF), margin factor (MF), waveform factor (WF) and kurtosis (Kur). For each discrete sequence of IC curves, the exact formulae for the dimensionless features are given as
\begin{equation}
	\begin{aligned}
		CF&=\frac{a_p}{a_{r m s}}=\frac{\max \left|a_i\right|}{\sqrt{\frac{1}{N} \sum_{i=1}^N a_i^2}}  \\
		PF&=\frac{a_p}{a^{\prime}}=\frac{\max \left|a_i\right|}{\frac{1}{N} \sum_{i=1}^N\left|a_i\right|}  \\
		MF&=\frac{a_p}{a_r}=\frac{\max \left|a_i\right|}{\left(\frac{1}{N} \sum_{i=1}^N \sqrt{\left|a_i\right|}\right)^2}  \\
		Kur&=\frac{1}{N} \sum_{i=1}^N \frac{a_i^4}{\left(\frac{1}{N} \sum_{i=1}^N a_i^2\right)^2}-3
	\end{aligned}
\end{equation}
where, $a_i$ is each sample individual; $a_p$ is the peak of the absolute value in the sequential sample; $a_{rms}$ is the root mean square of the sample, and $a_r$ is the square root mean square. The more technical details of extracting HI based on ICA used in this paper can be found in our previous work \cite{jia2025incremental}.
\subsection{Dataset}
\subsubsection{Oxford battery dataset}\label{subsec:oxford}
\begin{figure}[!htbp]
	\centering
	\includegraphics[width=.5\textwidth]{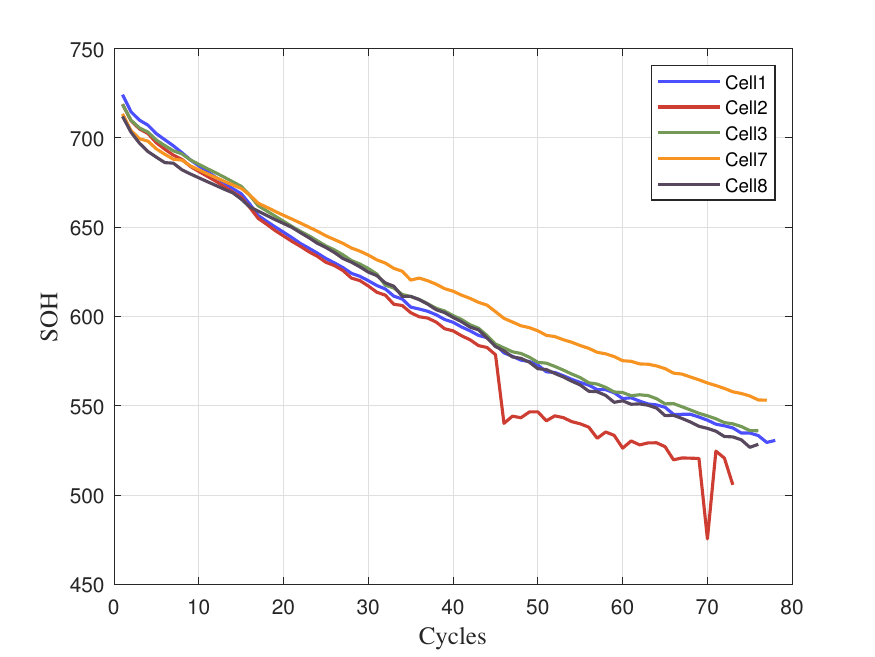}
	\caption{Capacity degradation curve of Oxford battery dataset.}
	\label{figur:f1}
\end{figure}

\begin{figure}[!htbp]
\centering
\includegraphics[width=.5\textwidth]{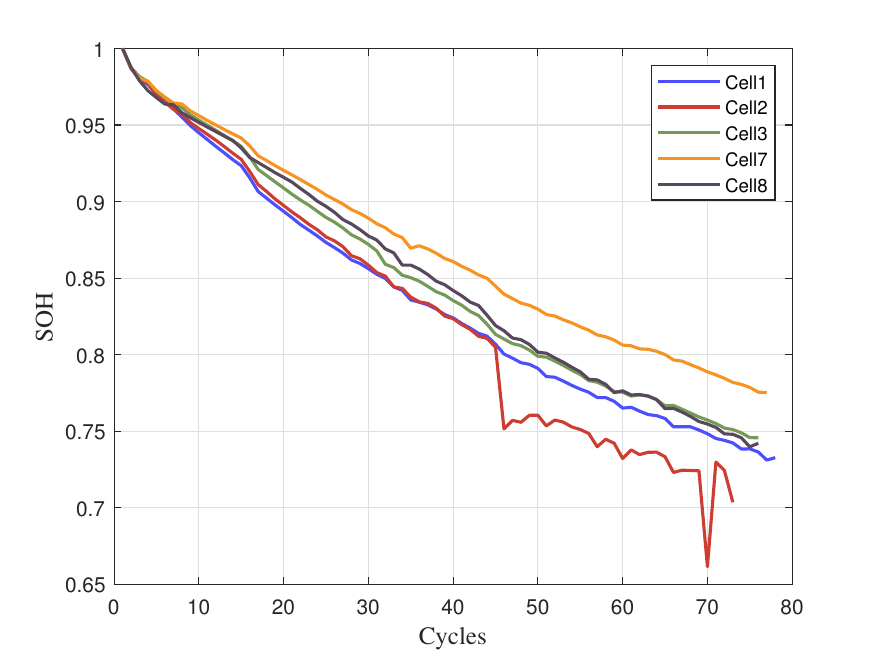}
\caption{SOH curves of Oxford battery dataset.}
\label{figur:f2}
\end{figure}

The Oxford dataset \cite{birkl2015parametric} contains eight Kokam pouch batteries rated at 740 mAh with graphite as the negative material as well as lithium cobaltate and lithium nickel cobaltate as the positive materials. The aging tests of these batteries can be categorized into capacity calibration tests and drive cycle tests. Specifically, the driving cycle conditions of a real EV are simulated as the load current for battery discharge, and then a complete test is realized at a 2 C charge rate. After every 100 cycles, charging/discharging at a small C rate was then performed to obtain the current available capacity. External parameters such as terminal voltage, surface temperature and load current were sampled at 1 s intervals by a Bio-Logic MPG-205 battery tester at a constant ambient temperature of 40 °C \cite{christoph2017diagnosis}. Five aging data, numbered Cell1, Cell2, Cell3, Cell7 and Cell8, were used in this paper to validate the proposed SSA-BiGRU model, as shown in Fig. \ref{figur:f1}. According to the definition of SOH in subsection \ref{subsec:soh}, the SOH curves of the selected five cells are shown in Fig. \ref{figur:f2}.

\begin{figure}[!htbp]
	\centering
	\includegraphics[width=.5\textwidth]{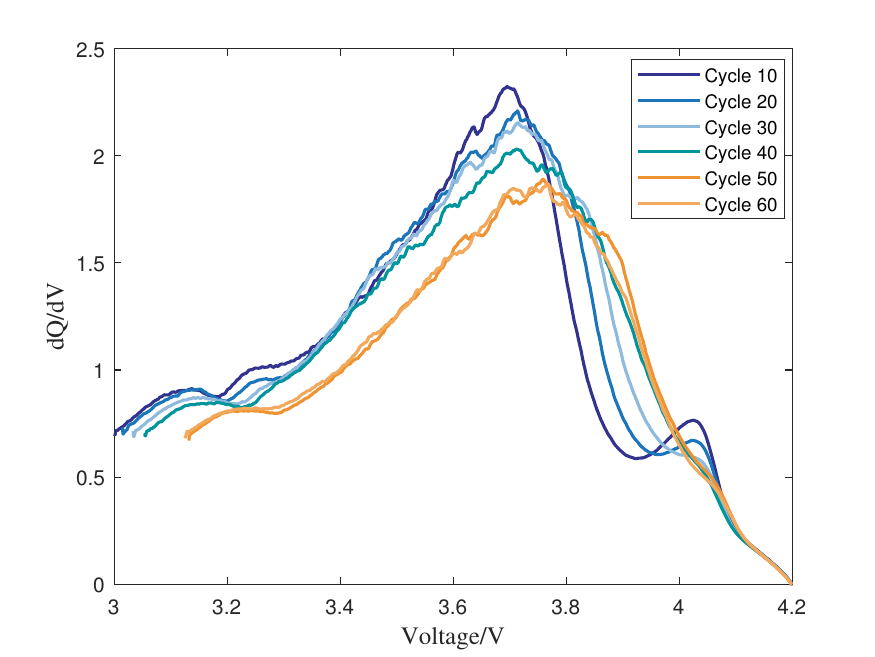}
	\caption{Partial charge IC curve of battery Cell2.}
	\label{figur:f3}
\end{figure}

The HI of LiBs comes from its own parameters obtained through monitoring and secondary analysis during operation, which can effectively characterize the degradation state of the battery in terms of data. ICA can reflect the internal changes of the battery to a certain extent, and the source of data is simpler and easier to obtain than the capacity or internal resistance. Therefore, obtaining effective and feasible HI through ICA is very important for SOH prediction of LiBs.The Oxford battery dataset simulates the charging and discharging under real driving conditions, while in reality, the complexity and uncertainty of the conditions lead to slightly different discharging conditions of the batteries. Thus, in this paper, IC curves under charging conditions are selected for the Oxford battery dataset to ensure smoothness and ease of feature extraction. Taking the battery Cell2  as an example, the IC part of the curve for its constant-current charging stage is shown in Fig. \ref{figur:f3} (plotted at intervals of 10 cycles starting from the 10th cycle). From Fig. \ref{figur:f3}, the peaks of the two peaks possessed by the IC curve decreased until they disappeared as the aging of the battery deepened with the cycling cycle, e.g., at the 10th cycle, two more distinct peaks existed near the voltages of 4.0 V and 3.65 V, whereas at the 60th cycle, the IC curve only had one peak near 3.67 V, which indicates that the battery ends the constant current charging process much earlier in the charging phase.

Using the HI extraction method in subsection \ref{subsec:ICA_HI} and our previous work \cite{jia2025incremental}, based on the IC curves extracted from the Oxford battery dataset, margin factor, area, peak, waveform factor, pulse factor and kurtosis in total 6 HIs can be obtained. Due to the large volatility and noise of the extracted HIs, we firstly perform a normalization process to deflate them to the interval of $\left[0, 1\right]$, and then perform singular value decomposition (SVD) \cite{zhao2017novel} denoising process for each normalized HI, which is denoted as SVD-HI. Finally, Spearman correlation analysis was performed between the denoised data and the SOH degradation curves of the battery, which can be calculated as
\begin{equation}\label{eq:correction}
	\text { Spearman }=\frac{\sum_i\left(x_i-\bar{x}\right)\left(y_i-\bar{y}\right)}{\sqrt{\sum_i\left(x_i-\bar{x}\right)^2} \sqrt{\sum_i\left(y_i-\bar{y}\right)^2}}
\end{equation}
where, $x_i$ and $y_i$ are the sample individuals; $\bar{x}$ and $\bar{y}$ are the corresponding average values.

\begin{figure}[!htbp]
	\centering
	\includegraphics[width=.9\textwidth]{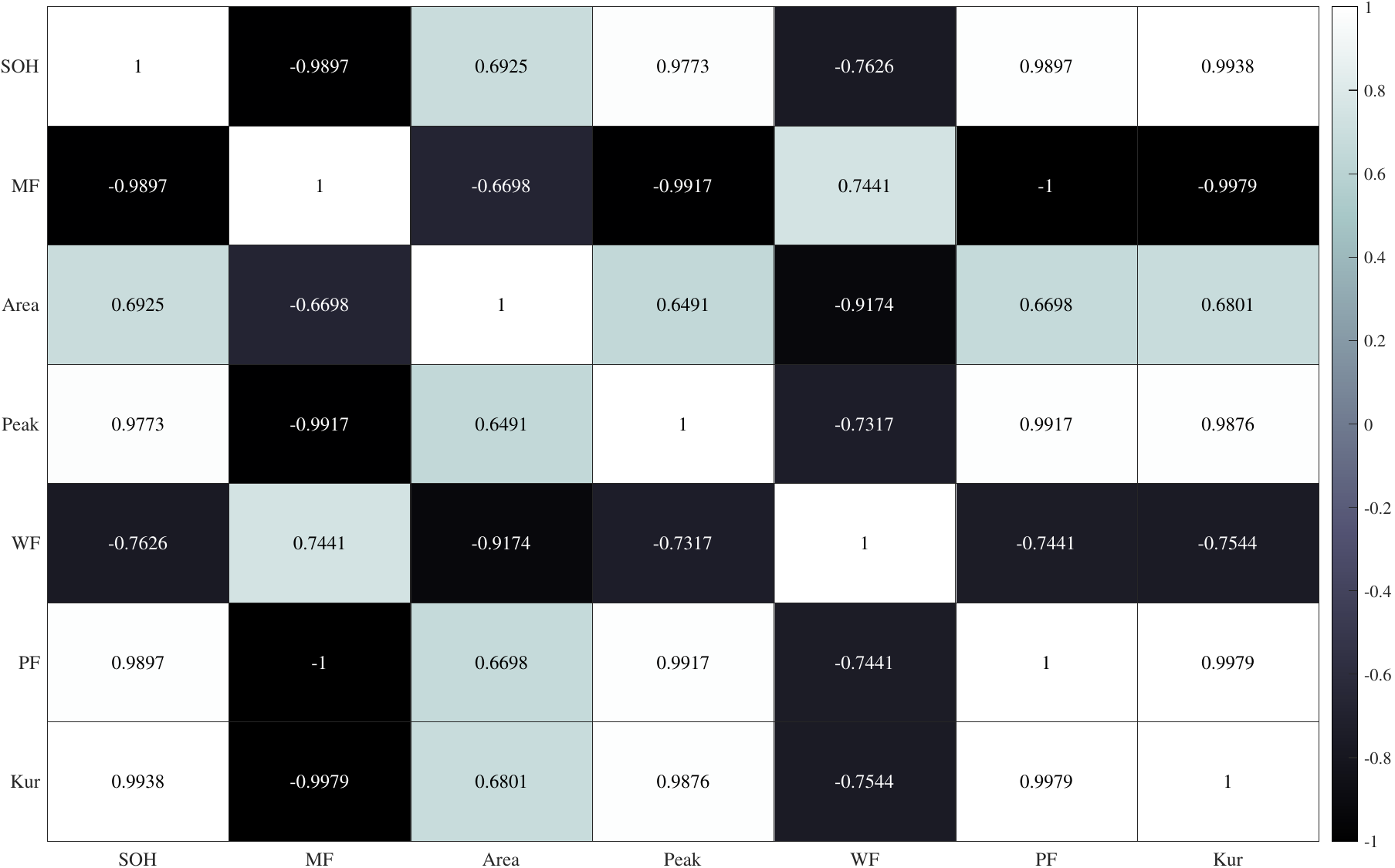}
	\caption{Thermogram of SVD-HI vs. SOH Spearman correlation for battery Cell2.}
	\label{figur:f4}
\end{figure}

Taking battery Cell2 as an example, its correlation heat map is shown in Fig. \ref{figur:f4}, from which for SVD-HI, the higher the positive or negative correlation with SOH, the closer the color embodied in the heat map is to white or black. Therefore, based on the positive and negative correlations, the margin factor, pulse factor, and kurtosis are initially selected as HIs for further use in this paper. Considering the specific model construction and input selection, specific ablation experiments will be shown in Section \ref{sec:experment}.

\subsubsection{Real road-driven EV charging dataset}\label{subsubsec:real_ev}
The real road-driven EV dataset used in this paper is derived from the battery pack charging data of 20 commercial EVs used by Deng et al. in \cite{deng2023prognostics}, with a time span of two years (about 29 months) of vehicle operation. The vehicles in the dataset are the BAIC EU500, equipped with the NCM 145 Ah battery type manufactured by Ningde Times and with a consistent battery system, which are numbered as Vehicle 1 (V1), Vehicle 2 (V2), $\dots$, Vehicle 20 (V20). The data were collected through the charging unit, which receives battery charging data during the charging process via control area network (CAN) communication. Charging data were recorded at a frequency of 8 s. The main items of the battery charging data and their resolution are summarized in Table \ref{Tab:t1}. The resolution of the real vehicle data is lower than that of the laboratory test data due to limitations in data transmission.

\begin{table}[!htbp]
	\caption{Battery charging data items}
	\label{Tab:t1}
	\centering
	\begin{tabular}{ccc}
		\toprule
		\multicolumn{1}{c}{Items}    & Units & \multicolumn{1}{c}{Resolutions} \\
		\midrule
		Current                     & A    & -                              \\
		Voltage                     & V    & 0.1 V                          \\
		SOC                         & -    & 0.1                            \\
		Maximum Battery Voltage     & V    & 0.001                          \\
		Minimum Battery Voltage     & V    & 0.001                          \\
		Maximum Battery Temperature & ℃    & 1 ℃                            \\
		Minimum Battery Temperature & ℃    & 1 ℃                           \\
		\bottomrule
	\end{tabular}
\end{table}

\begin{figure}[!htbp]
	\centering
	\includegraphics[width=1\textwidth]{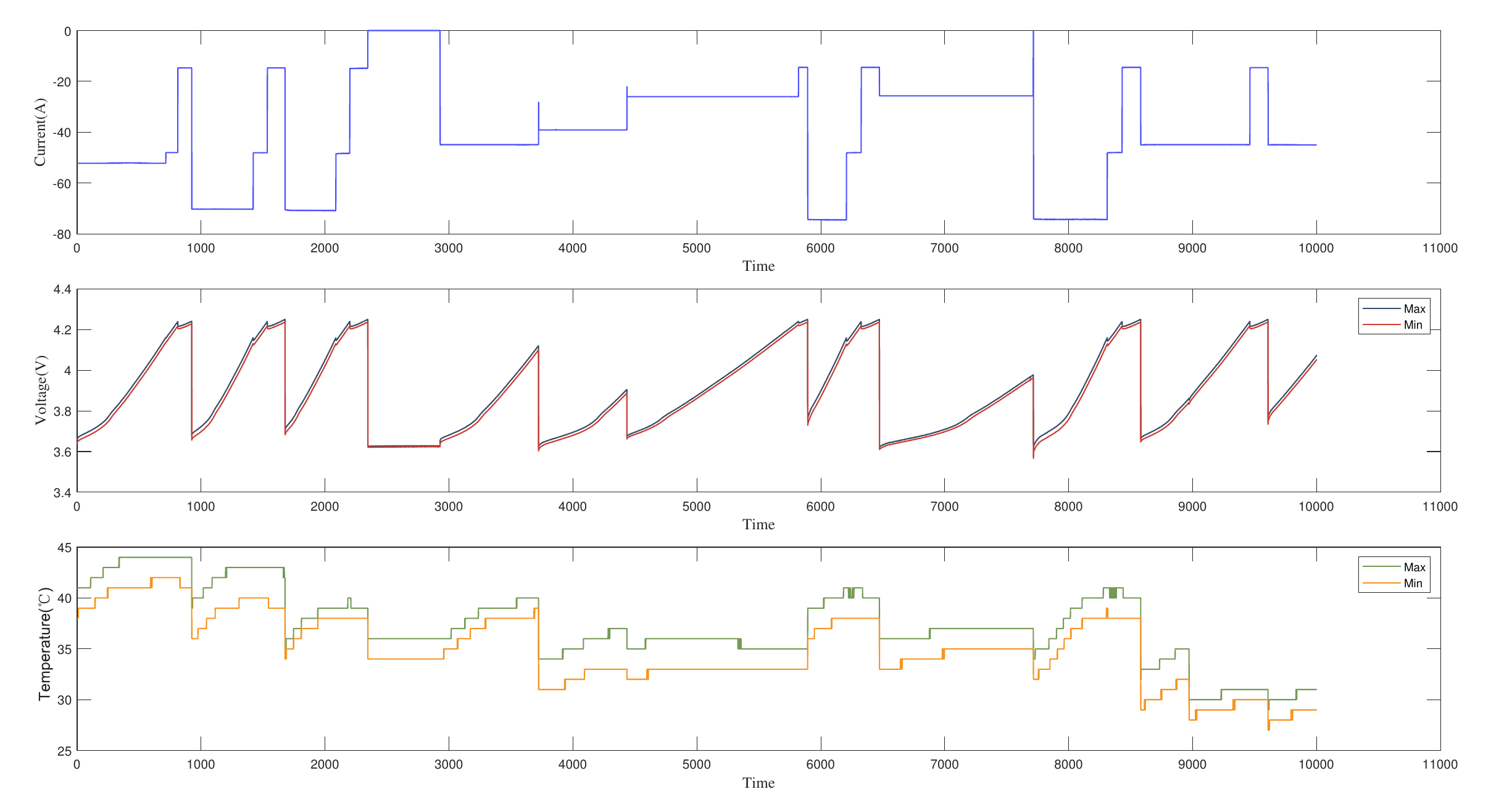}
	\caption{Charging curves of V1.}
	\label{figur:f5}
\end{figure}

The dataset uses a variant of the ampere-time integration method to calculate battery capacity, i.e.,
\begin{equation}\label{eq:capacity}
	C=\frac{-\int_{t_1}^{t_2} I(t) \Delta t}{S O C_{t_2}-S O C_{t_1}}
\end{equation}
where, $\Delta t$ is the fixed sampling interval; $I(t)$ is the battery current (negative) for the charging process; $t_1$ and $t_2$ are the charging start time and end time, respectively.
Taking V1 as an example, the spliced data of the charging process of its first 10,000 measurement points is shown in Fig. \ref{figur:f5}, from which the charging strategy of these vehicles is multi-stage constant current charging, and the current value of each stage is determined by the temperature of the battery.

\begin{figure}[!htbp]
	\centering
	\includegraphics[width=1\textwidth]{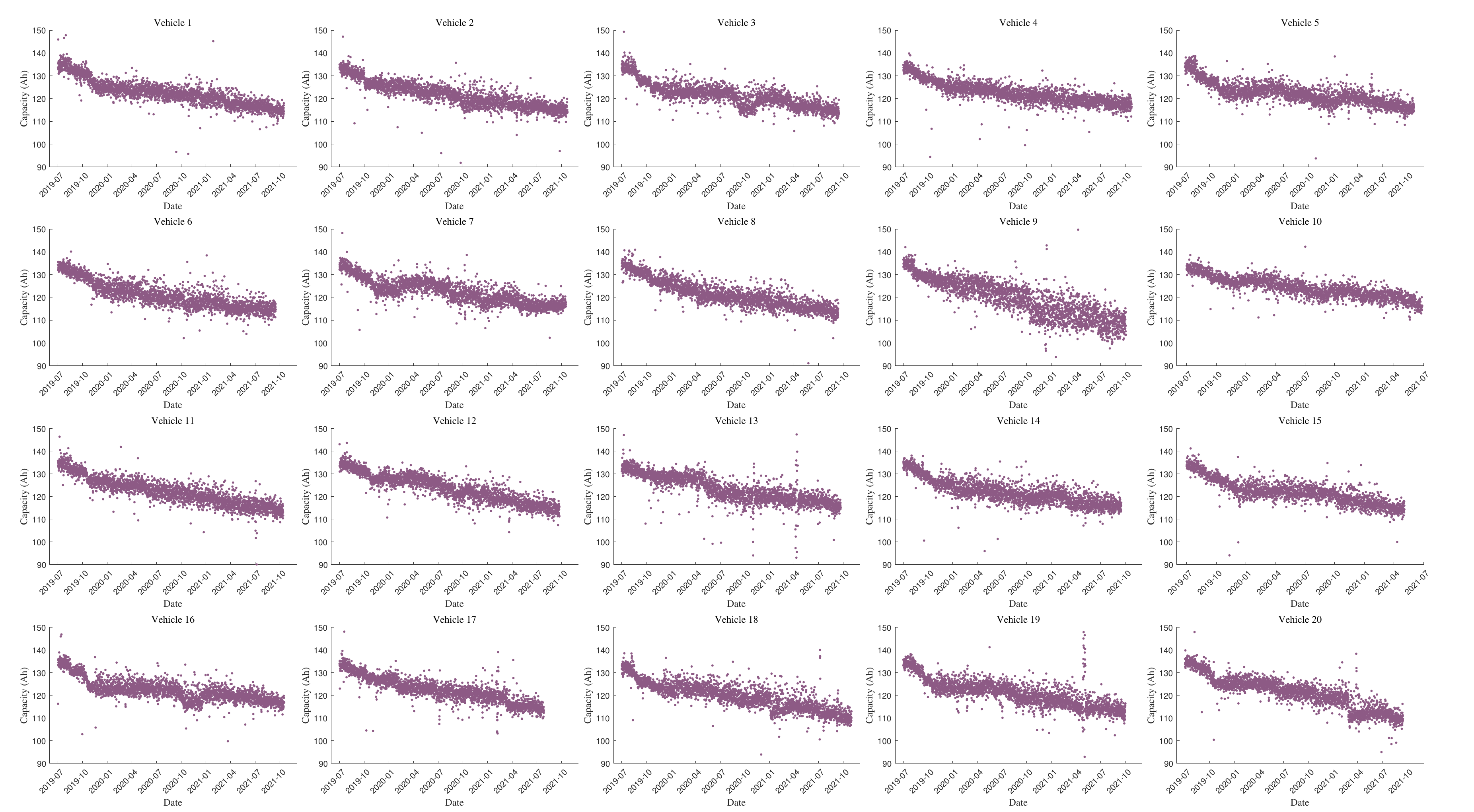}
	\caption{Raw charging data of 20 EVs.}
	\label{figur:f6}
\end{figure}

\begin{figure}[!htbp]
	\centering
	\includegraphics[width=1\textwidth]{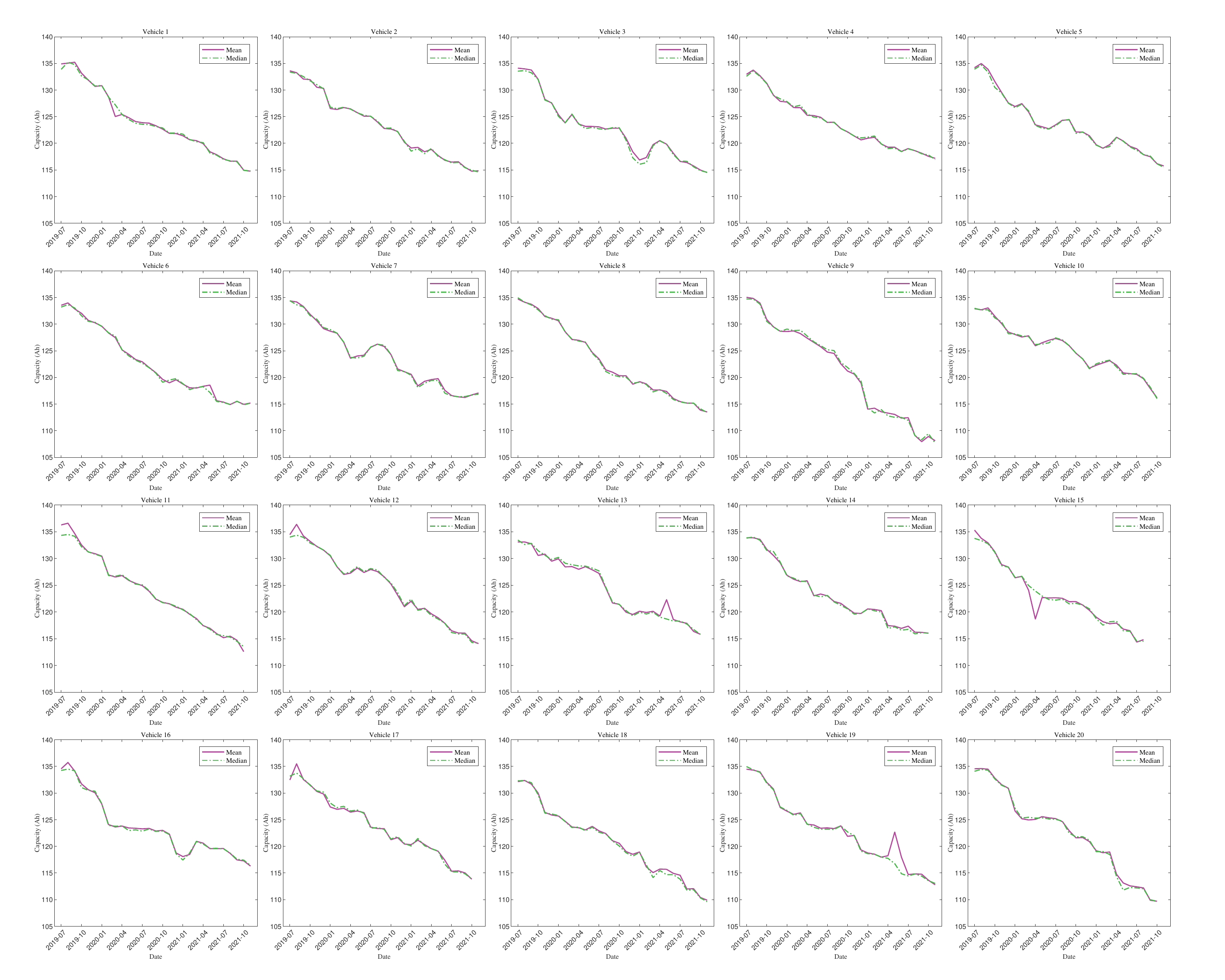}
	\caption{Monthly average and median charging of 20EVs.}
	\label{figur:f7}
\end{figure}

The raw charging data for the 20 EVs is shown in Fig. \ref{figur:f6} and it can be seen that there are some outliers in the raw data and a large number of points that fluctuate within the same month. For each month, an average of 90 data points can be obtained, and the mean and median are taken for each month's data respectively. The results obtained are shown in Fig. \ref{figur:f7}. As can be seen from  Fig. \ref{figur:f7}, the mean is almost equal to the median, which indicates a symmetrical distribution of the calculated capacity points within a month, and both can be used to effectively represent the degradation state of the battery system. There are localised differences between the median and the mean for individual vehicle batteries, such as V12, V15 and V19. Moreover, it can be observed that there are multiple localised capacity rebound processes in the battery data for these EVs, which may be caused by the factors such as prolonged periods of non-use and temperature changes. Considering the statistical significance of the mean and median, the median battery degradation was chosen to represent the degradation status in this paper in order to reduce the interference of the possible extreme values in the data for each month.

\subsection{Evaluation indicators}\label{subsec:indicators}
In this paper, three error evaluation metrics are used to quantitatively assess the output of the prediction model, namely root mean square error (RMSE), mean absolute error (MAE) and mean absolute percentage error (MAPE ), which are defined as
\begin{equation}\label{eq:rmse}
	RMSE=\sqrt{\frac{1}{N} \sum_{i=1}^N\left(x_i-\hat{x}_i\right)^2}
\end{equation}
\begin{equation}\label{eq:mae}
	M A E=\frac{1}{N} \sum_{i=1}^N\left|x_i-\hat{x}_i\right|
\end{equation}
\begin{equation}\label{eq:mape}
	M A P E=\frac{1}{N} \sum_{i=1}^N\left|\frac{x_i-\hat{x}_i}{x_i}\right| \times 100 \%
\end{equation}
where, $x_i$ and $\hat{x}_i$ indicates the true value and predicted value of SOH, respectively. The smaller the value of the above evaluation indicators, the higher the prediction accuracy is indicated.

\section{SSA-BiGRU-based SOH prediction model and prediction process}\label{sec:SSA-DBiGRU}
\subsection{BiGRU and SSA}
\subsubsection{BiGRU}
\begin{figure}[!htbp]
	\centering
	\includegraphics[width=.35\textwidth]{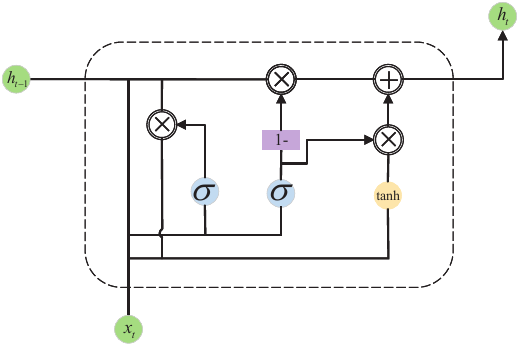}
	\caption{Structure of GRU.}
	\label{figur:f8}
\end{figure}

In 1986, David Rumelhart innovatively introduced the concept of RNN \cite{rumelhart1986learning}, which places more emphasis on the correlation between data and determines the output of the current state based on past memories than traditional neural networks. However, when there are long-term dependencies in the sequence, the RNN fails severely and suffers from gradient explosion and gradient vanishing problems \cite{chemali2017long}. LSTM \cite{hochreiter1997long} and GRU \cite{cho2014learning}, as variant structures of RNN, address the limitations of RNN by introducing a gating mechanism, which enables it to selectively remember key information. LSTM has three gates: forget gate, input gate, and output gate. Different from LSTM, GRU combines the input gate and forget gate into a single update gate, which controls the flow of information and due to this simplification. GRU requires fewer trainable parameters and hence it has a simpler architecture and requires less computation time during training as compared to LSTM. The  structure of GRU is shown in Fig. \ref{figur:f8}.

When dealing with sequential tasks, GRU can take the past hidden state $h_{t-1}$ as input and output the new hidden state $h_t$. The update gate and reset gate are denoted by $U_t$ and $R_t$, respectively.
\begin{itemize}
	\item Update gate: The update gate is intended to define the information to be stored through a linear computation and non-linear process, which is used to control the flow of information transferred from the previous hidden state $h_{t-1}$ to the current hidden state $h_t$ and to determine the processing of long-term dependencies of the time series data. The value of the update gate determines the amount of information stored at the previous moment, and this amount of information is proportional to the value of the update gate. The formula of update gate is
		$U_t=\sigma\left(W_U \cdot\left[x_t, h_{t-1}\right]+b_U\right)$,
	in which $x_t$ is the current input of GRU, $h_{t-1}$ represents the previous output, $W_U$ and $b_U$ are the weight and bias of update gate, respectively, and $\sigma$ represents the Sigmod activation function.
	\item Reset gate: The role of the reset gate $R_t$ is to determine whether or not the value of the storage unit is forgotten and how much of the past information is forgotten. By controlling the degree to which $h_{t-1}$ and $x_t$ are combined, the reset gate is able to determine which past information should be ignored and which should be retained to help the model better capture short-term dynamics in the sequence. The value of the reset gate lies $\left[0,1\right]$, the closer it is to $0$, the GRU will take less account of past information and tend to ignore previous states, while when its value is close to one, the GRU will take full account of the information stored in the past, making the old state $h_{t-1}$ more involved in the construction of the new state $h_t$. The formula of reset gate is
		$R_t=\sigma\left(W_R \cdot\left[x_t, h_{t-1}\right]+b_R\right)$,
	where $W_R$ and $b_R$ represent weights and biases, respectively. The candidate output state $\tilde{h}_t$ will utilize the past relevant information stored by the reset gate, i.e., 
		$\tilde{h}_t={\rm{tanh}}\left(W_h \cdot\left[x_t, R_t, h_{t-1}\right]+b_h\right)$,
	where $W_h$ and $b_h$ represent the weights and biases of the candidate state $\tilde{h}_t$, respectively, $\rm{tanh}$ is a nonlinear activation function for deflating the information into the interval $\left[-1,+1\right]$.
	\item New hidden state $h_t$: The output of GRU consists of output information from the previous moment and current moment can be described as $h_t=\left(1-U_t\right)\cdot h_{t-1}+\tilde{h}_t\cdot U_t$, where the strength of the dependence between $h_t$ and $\tilde{h}_t$ is positively correlated with the size of $U_t$.
\end{itemize}

\begin{figure}[!htbp]
	\centering
	\includegraphics[width=.75\textwidth]{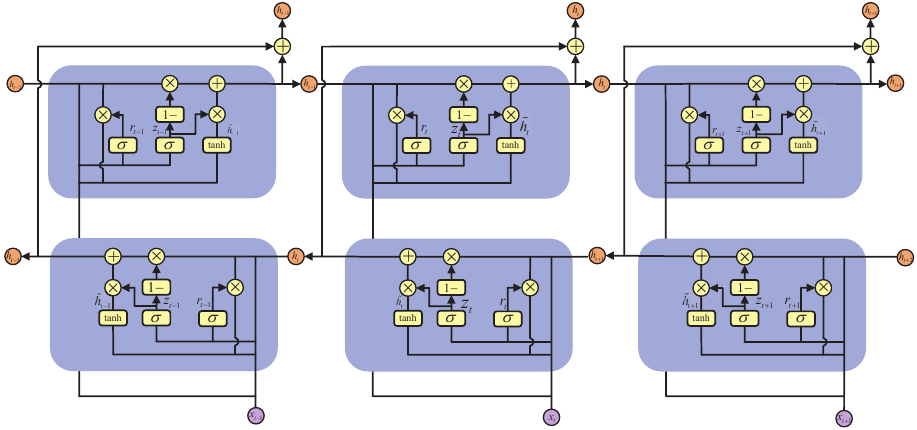}
	\caption{Schematic diagram of BiGRU.}
	\label{figur:f9}
\end{figure}

When dealing with time series, a unidirectional GRU can only transmit information unidirectionally, which means that it can only obtain information at a point in time in the past, while ignoring information at a point in time in the future. To solve this problem, this paper uses the dual-module BiGRU model, which contains two BiGRU modules. The BiGRU consists of two layers of GRUs that can use both past and future information, as shown in Fig. \ref{figur:f9}. BiGRU extracts information from continuous data simultaneously through interconnected hidden layers, by processing the sequences in forward and reverse direction respectively, and finally generates the output $Y_t$ of the whole network from the forward and reverse outputs, which is described as
\begin{equation}
	\begin{aligned}
		\overrightarrow{H}&=\left\{\overrightarrow{h}_1, \overrightarrow{h}_2, \overrightarrow{h}_3, \cdots, \overrightarrow{h}_i, \overrightarrow{h}_{i+1}, \cdots, \overrightarrow{h}_t\right\} \\
		\overleftarrow{H}&=\left\{\overleftarrow{h}_1, \overleftarrow{h}_2, \overleftarrow{h}_3, \cdots, \overleftarrow{h}_i, \overleftarrow{h}_{i+1}, \cdots, \overleftarrow{h}_t\right\} \\
		Y_t&=\left(\overrightarrow{H}, \overleftarrow{H}\right)
	\end{aligned}
\end{equation}
where, $\overrightarrow{H}$ and $\overleftarrow{H}$ are forward and reverse hidden state sequences, respectively.
\subsubsection{SSA}
Inspired by the predatory characteristics of sparrows, Xue et al proposed the group intelligence optimization algorithm SSA in 2020 \cite{xue2020novel}. SSA has the advantages of  global search performance and fast convergence rate. SSA is based on the sparrow's observation of its group members while foraging. According to the swarming characteristics that its species has, the sparrow observes other individuals in the group at the same time when it carries out foraging activities. Based on the foraging behavior of individuals, members of a sparrow group can be divided into producers and gleaners, with producers actively searching for food resources and gleaners acquiring food resources under the guidance of producers. In addition, birds have the flexibility to switch between producers and scavengers, i.e., in order to find food, sparrows usually employ both producer and scavenger strategies. Some individual sparrows compete with other sparrows for high-quality food resources to take their place. It is worth noting that the energy reserves of individuals play a key role when sparrows choose different foraging strategies: individuals with lower energy reserves tend to act as scavengers.

\begin{figure}[!htbp]
	\centering
	\includegraphics[width=.45\textwidth]{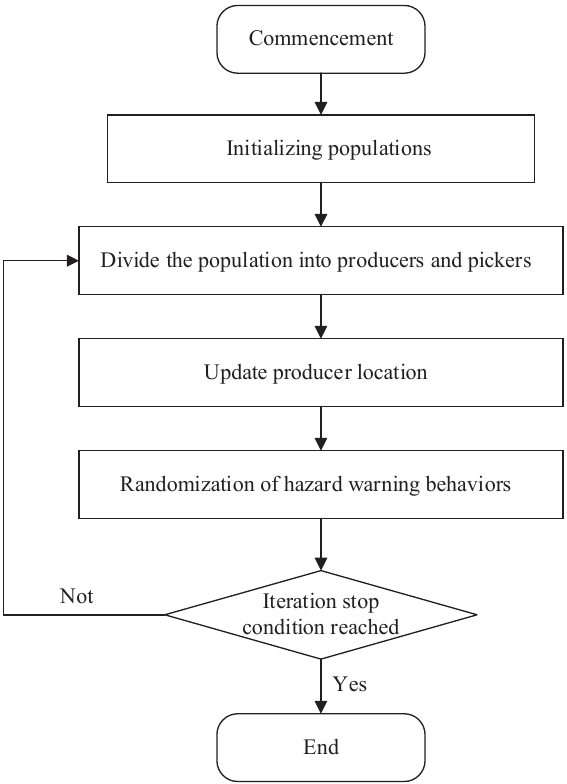}
	\caption{Solution process of SSA.}
	\label{figur:f10}
\end{figure}

SSA updates the location information of producers, pickers and warners by iteratively updating the location information of food producers, gleaners and warners based on individual fitness, which ultimately leads to the identification of the location information of the global optimizer. The solution process of SSA is shown in Fig. \ref{figur:f10}. Initially, the algorithm classifies the population based on the optimization variables and subsequently divides them into producers and gleaners. Then, it updates their locations based on the fitness values of the individuals. A portion of them is randomly selected as early warners and their positions are updated accordingly. The process iterates until the termination criteria are satisfied. If the criterion is not satisfied, the iteration continues; otherwise, the algorithm stops and provides the optimal solution.

In each iteration, the position of the producer is updated as
\begin{equation}\label{eq:ssa_producer}
	X_{i, j}^{t+1}=\left\{\begin{array}{lc}
		X_{i, j}^t \cdot e^{\frac{-i}{\alpha \cdot \rm{iter}_{\mathrm{max}}}} & \text { if } R_2<S T \\
		X_{i, j}^t+Q \cdot L & \text { if } R_2 \geq S T
	\end{array}\right.
\end{equation}
where, $t$ is the current iteration number; $X_{i,j}^t$ denotes the value of the $i$th sparrow individual in the $j$th dimension, $j=1,2,\cdots, D$; $\rm{iter}_{max}$ is the constant with the highest number of iterations, and $\alpha\in\left(0,1\right]$ is a random number; $R_2\in\left[0,1\right]$ and $ST\in\left[0.5,1\right]$ denote the warning value and the safety threshold, respectively; $Q$ is a random number obeying a normal distribution, and $L$ denotes a $1\times D$ matrix with each element being $1$.

The position of the picker is
\begin{equation}\label{eq:ssa_picker}
X_{i, j}^{t+1}=\left\{\begin{array}{lc}
	Q \cdot e^{\frac{X_{\text {worst }}^t-X_{i, j}^t}{i^2}} & \text { if } i>n / 2 \\
	X_P^{t+1}+\left|X_{i, j}^t-X_P^{t+1}\right| \cdot A^{+} \cdot L & \text { otherwise }
\end{array}\right.
\end{equation}
where, $Q$ is the best position occupied by the producer and $X_{\text {worst }}$ denotes the current global worst position; $A$ is the $1 \times D$ matrix in which each element is randomly assigned to be either $1$ or $-1$, and $A^{+}=A^T\left(A A^T\right)^{-1}$.

Additionally, the position of the warner was randomly generated within the sparrow population, which is given as
\begin{equation}\label{eq:ssa_warners}
X_{i, j}^{t+1}=\left\{\begin{array}{cl}
	X_{\text {Best }}^t+\beta \cdot\left|X_{i, j}^t-X_{\text {best }}^t\right| & \text { if } f_i>f_g \\
	X_{i, j}^t+K \cdot\left(\frac{\left|X_{i, j}^t-X_{\text {worst }}^t\right|}{\left(f_i-f_w\right)+\varepsilon}\right) & \text { if } f_i=f_g
\end{array}\right.
\end{equation}
where, $X_{\text {best }}$ is the current global optimal position, and $\beta$ is the control parameter for the step size (random numbers with mean 0 and variance 1 that follow a normal distribution); $K$ is a random number within $\left[-1,1\right]$; $f_i$ denotes the current fitness value of the sparrow, and $f_w$ and $f_g$ represent the current global worst fitness value and best fitness value, respectively; $\varepsilon$ is the smallest constant to prevent a denominator of $0$; $X_{\text {best }}$ represents the center of the population, around which it is safe; $K$ represents both the transfer direction of sparrows and the control coefficient of step size.
\subsection{SSA-BiGRU-based SOH prediction model for LiBs}
In this subsection, a fusion model combining full-parameter-domain SSA and dual-module BiGRU is proposed. The dual-module BiGRU contains two BiGRU modules, which adopt a two-module architecture for extracting time series features in both directions, and SSA is used to find the optimal combinations of the independent hyperparameters of the dual-module BiGRU model. The fusion model combines the advantages of BiGRU and SSA for SOH prediction of LiBs in EVs.

\begin{figure}[!htbp]
	\centering
	\includegraphics[width=1\textwidth]{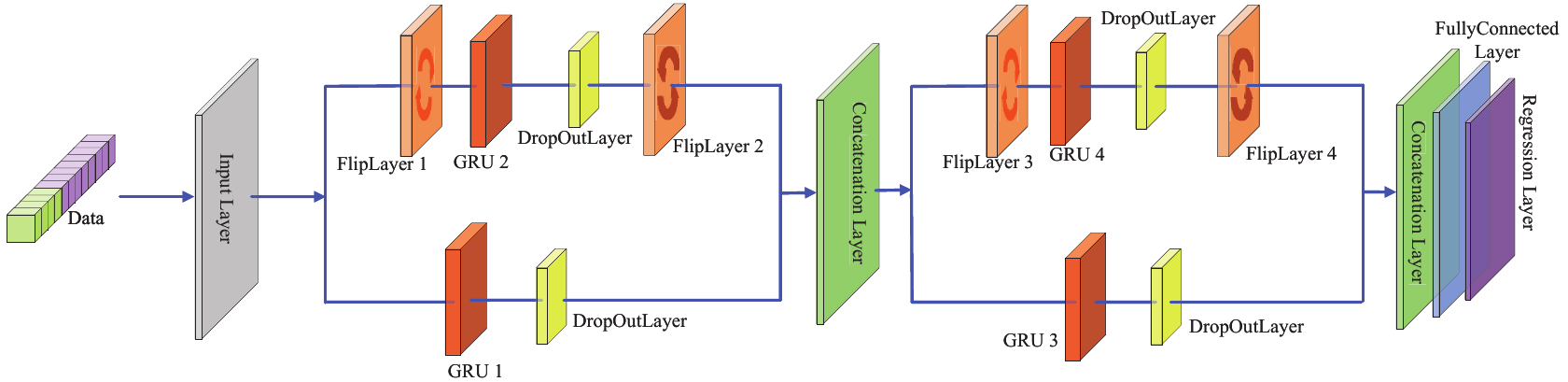}
	\caption{Schematic diagram of the network layer structure of the dual-module BiGRU model.}
	\label{figur:f11}
\end{figure}

The overall architecture and network layer structure of the dual-module BiGRU model used in this paper is shown in Fig. \ref{figur:f11}, where the main body of the model consists of two BiGRU modules, each of which is divided into forward and backward propagating two-layered GRUs, with serial connections between the modules. The forward GRU consists of GRU layer and DropOut layer, while the backward GRU consists of Flip layer, GRU layer, DropOut layer and Flip layer.

The model structure of the dual-module BiGRU model is described in detail as follows.
\begin{itemize}
	\item \textbf{Input layer}: A portion of the HI extracted based on the IC curve is divided as input to train the model;
	\item \textbf{Dual-module BiGRU}: The forward GRU network is responsible for capturing the information of historical data to current data in the sequence, while the information of future data to current data in the sequence is taken care of by the backward GRU. For the backward GRU, the sequence is first inverted by the Filp layer, and the result is outputted by the Filp layer after being learnt by the GRU. Compared to the conventional BiGRU, the use of two BiGRU modules (i.e., dual-module BiGRU) increases the depth of the neural network and strengthens the learning ability of the network;
	\item \textbf{DropOut layer}: Reduce overfitting of the model and improves the generalization of the model;
	\item \textbf{Output layer}: The output of the dual-module BiGRU module is mapped to the SOH prediction using a fully connected layer, and finally the loss function is computed using a regression layer.
\end{itemize}

The hyperparameters in the dual-module BiGRU model are optimized independently using SSA, including the number of units in each GRU layer, the dropout rate in the DropOut layer, and other network parameters in the Adam optimizer. The number of populations was set to $6$ and the maximum number of iterations was set to $10$. The specific parameters and optimization ranges are shown in Table \ref{Tab:t2}.

\begin{table}[!htbp]
	\caption{SSA-BiGRU parameters and their optimization range}
	\label{Tab:t2}
	\centering
\begin{tabular}{cc}
	\toprule Parameters &  Optimization Ranges \\
	\midrule
	Number of GRU1 cells in the hidden layer & $[25,200]$ \\
	Number of GRU2 cells in the hidden layer& $[25,200]$ \\
	Number of GRU3 cells in the hidden layer& $[25,200]$ \\
	Number of GRU4 cells in the hidden layer& $[25,200]$ \\
	 MaxEpochs & $[150,700]$ \\
	 Learning Rate & $[0.005,0.015]$ \\
	 LearnRateDropPeriod & $0.7 *$MaxEpochs \\
	 LearnRateDropFactor & $0.01$\\
	 Minimum batch size & $[1,20]$ \\
	 Dropout Rate of DropOut Layer 1& $[0.002,0.2]$ \\
 	 Dropout Rate of DropOut Layer 2& $[0.002,0.2]$ \\
	 Dropout Rate of DropOut Layer 3& $[0.002,0.2]$ \\
	 Dropout Rate of DropOut Layer 4& $[0.002,0.2]$\\
	\bottomrule
\end{tabular}
\end{table}

\subsection{SSA-BiGRU-based SOH prediction flow for LiBs}
\begin{figure}[!htbp]
	\centering
	\includegraphics[width=1\textwidth]{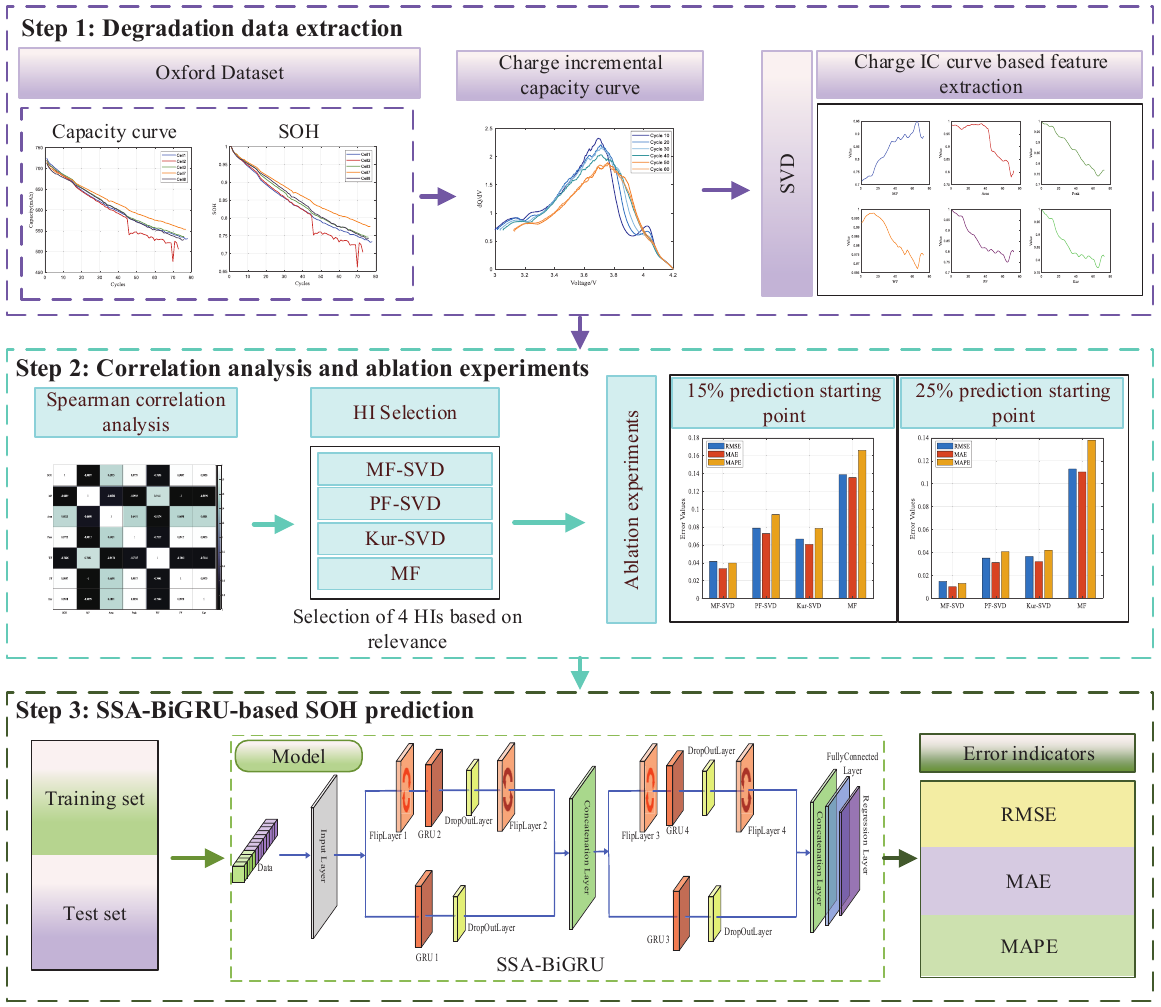}
	\caption{SOH prediction flowchart based on SSA-BiGRU model.}
	\label{figur:f12}
\end{figure}

The flowchart of SOH prediction for LiBs based on the SSA-BiGRU model is shown in Fig. \ref{figur:f12}, from which the SOH prediction consists of three steps.
\begin{enumerate}[\textbf{Step} 1]
	\item Degradation data extraction. The dimensionless features are extracted from the IC curves in the charging phase and smoothed with the help of SVD to reduce the effect of noise on model training.
	\item Correlation analysis and ablation experiments. Through Spearman correlation analysis, three dimensionless features with the highest correlation are selected. In order to verify the HI with the best training effect, taking Cell2 cell as an example, one can take these three HIs and the margin factor without SVD treatment as training data respectively, and performs SOH prediction through the dual-module BiGRU model, and the best HI is screened out according to the prediction error.
	\item SSA-BiGRU-based SOH prediction. Based on the work in \textbf{Step 1} and \textbf{Step 2}, the best combination of HI and prediction model was successfully selected, followed by experiments on four sets of battery data except Cell2. The dual-module BiGRU model is used to learn the bidirectional information of the data, the model is fully parameter optimized using SSA to enhance its generalization performance, and finally the prediction results are compared with the multi-group fusion model to evaluate the performance of the prediction model in all aspects.
\end{enumerate}

\section{SOH prediction experiment and result analysis based on SSA-BiGRU model}\label{sec:experment}
In this section, the Oxford battery dataset and the real road-driven EV charging dataset are selected as the research objects, which are used to verify the effectiveness of the proposed SSA-BiGRU model. The comparison experiment of HI selection and the ablation experiment of SSA-BiGRU model are carried out based on the data of battery Cell2 in Oxford battery dataset. On this basis, the comparison experiments of different prediction starting points and different models are carried out on the remaining four sets of battery data in Oxford battery dataset. Finally, the generalization of the SSA-BiGRU model is verified on the real road-driven EV charging dataset.
\subsection{HI comparison experiment}\label{subsec:HI}
In this subsection, the experiments of SOH prediction are performed on battery Cell2 based on each of the three HIs (margin factor, pulse factor, and kurtosis) after SVD denoising and Spearman analysis shown in subsection \ref{subsec:oxford}. In addition, the margin factor without SVD denoising is additionally attached as a comparison to verify the necessity of SVD denoising. Since both the HI sequence and the actual SOH sequence are one-dimensional and of equal length, the first 15\% and 25\% of the sequential data of the HI sequence are selected as the training set, and the experiments are performed on the data after 15\% and 25\% of the SOH sequence, respectively. The dual-module BiGRU model without SSA optimization was chosen for the prediction model, and the parameter settings are shown in Table \ref{Tab:t3}.
\begin{table}[!htbp]
	\caption{BiGRU parameter settings}
	\label{Tab:t3}
	\centering
	\begin{tabular}{cc}
		\toprule Parameters &  Values \\
		\midrule
		Number of neurons in the hidden layer & 128 \\
		MaxEpochs& 500 \\
		Learning Rate& 0.01 \\
		LearnRateDropPeriod& 350\\
		LearnRateDropFactor & 0.01 \\
		Minimum batch size& 16 \\
		DropOut Layer Discard Rate & 0.02\\
		\bottomrule
	\end{tabular}
\end{table}

\begin{figure}[!htbp]
	\centering
	\subfigure[Prediction starting point at 15\%.]{\label{figur:f13a}	\includegraphics[width=0.45\textwidth]{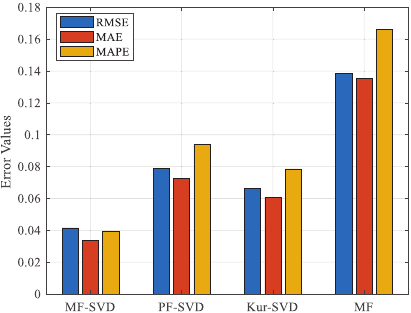}}
	\subfigure[Prediction starting point at 25\%.]{\label{figur:f13b}	\includegraphics[width=0.45\textwidth]{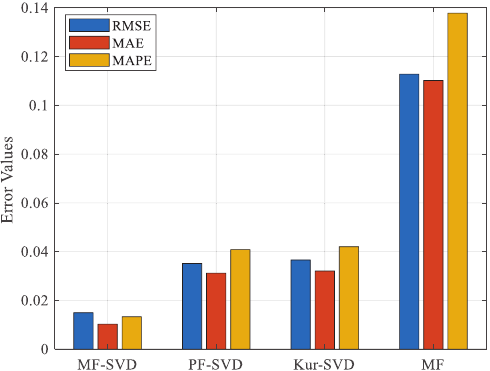}}
	\caption{Errors for battery Cell2 at different prediction starting points.}
	\label{figur:comp_different_start_point}
\end{figure}

For the experimental results, RMSE, MAE and MAPE are chosen in this subsection as comparisons, and the error histograms for the two sets of prediction starting points are shown in Fig. \ref{figur:f13a} and Fig. \ref{figur:f13b}, respectively. In order to keep the order of magnitude consistent for graphical visualisation, MAPE is uniformly reduced by two orders of magnitude in this subsection.

As can be seen from Fig. \ref{figur:comp_different_start_point}, MF-SVD gets the best error in terms of prediction effectiveness, which is lower than PF-SVD and Kur-SVD across the board. Moreover, comparing to MF without SVD denoising, the error values, in terms of RMSE, for example, decrease by 70\% and 86.76\%, respectively. For MF-SVD, its RMSE decreased by 64.07\% after 10\% increase in training data. Therefore, based on the two sets of experiments, MF-SVD is chosen as the final HI in this paper.

\subsection{Model ablation experiments on SSA-BiGRU model}\label{subsec:model_ablation}
\begin{figure}[!htbp]
	\centering
	\subfigure[Prediction starting point at 5\%.]{\label{figur:f14a}	\includegraphics[width=0.32\textwidth]{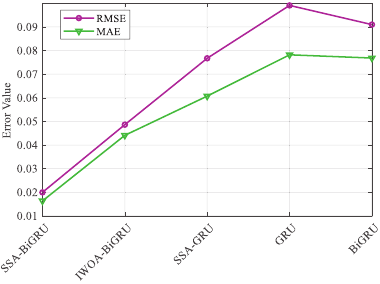}}
	\subfigure[Prediction starting point at 15\%.]{\label{figur:f14b}	\includegraphics[width=0.32\textwidth]{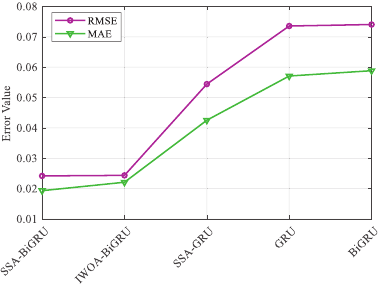}}
	\subfigure[Prediction starting point at 25\%.]{\label{figur:f14c}	\includegraphics[width=0.32\textwidth]{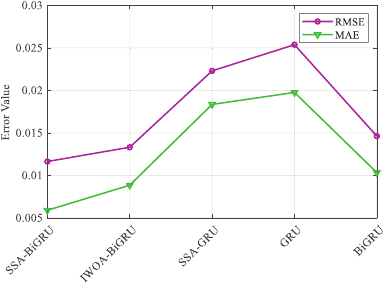}}
	\caption{Error of model ablation for battery Cell2.}
	\label{figur:ablation_models}
\end{figure}

Since SSA-BiGRU consists of two components and there is a single comparable component for each of them, we sets up a model ablation experiment on SSA-BiGRU based on the data of battery Cell2 and the HI selection strategy derived in subsection \ref{subsec:HI}, comparing the models of SSA-GRU, BiGRU and GRU, where a two-layer architecture is adopted for the part involving the GRU and the consistent in depth with the model proposed in this paper. Considering that different optimization algorithms can also have different optimization results for the same model, the improved whale optimization algorithm (IWOA) \cite{IWOA2020} is chosen in this subsection as a comparison of SSA  in order to verify the effectiveness of SSA for the dual-module BiGRU model. Moreover, in addition to the 15\% and 25\% prediction starting points, an additional 5\%  prediction starting point is set in this subsection to further compare the different prediction models. The RMSE and MAE of the prediction results are shown in Fig. \ref{figur:ablation_models}.

According to Fig. \ref{figur:ablation_models}, it can be seen that among the three sets of experiments for different prediction starting points, both RMSE and MAE of the SSA-BiGRU model achieve the smallest error, and both are better than the BiGRU model without optimisation algorithm, and the RMSE was slightly weaker than that of the IWOA-BiGRU model only in Fig. \ref{figur:f14b}, which indicates that the optimal combination of SSA and dual-module BiGRU was achieved. In contrast, among the comparison models, the single GRU model has the worst prediction effect, followed by the SSA-BiGRU model, indicating that unidirectional GRU learning is weaker than BiGRU, and the errors of the two are only equal at 15\% prediction starting point. In a side-by-side comparison, most of the prediction errors achieve a decrease with more training data, and although the error of the SSA-BiGRU model fluctuates in the 5\% and 15\% prediction starting points, it decreases rapidly when the training data is increased to 25\%: RMSE decreases by almost two orders of magnitude, which is not the case for any other model except for the IWOA-BiGRU model.

\subsection{SOH prediction and comparisons of different models}\label{subsec:oxford_prediction}
\begin{figure}[!htbp]
	\centering
	\subfigure[Predictions starting point at 5\%.]{\label{figur:f15a}	\includegraphics[width=0.32\textwidth]{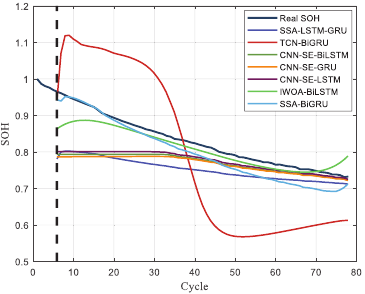}}
	\subfigure[Predictions starting point at 15\%.]{\label{figur:f15b}	\includegraphics[width=0.32\textwidth]{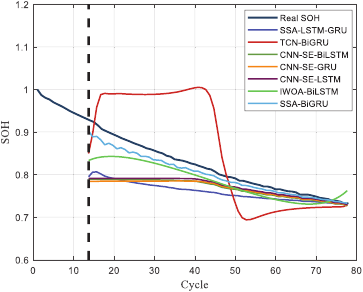}}
	\subfigure[Predictions starting point at 25\%.]{\label{figur:f15c}	\includegraphics[width=0.32\textwidth]{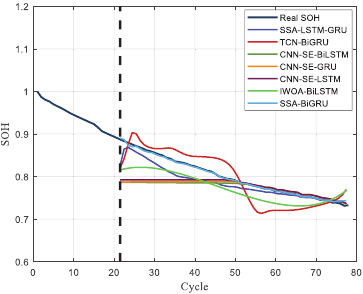}}
	\caption{Experiment results of model comparison for different prediction starting points for battery Cell1.}
	\label{figur:diff_models}
\end{figure}

In this subsection, the SSA-BiGRU model is used to predict the SOH at different prediction starting points for the remaining four sets of battery data in the Oxford battery dataset. In order to broadly compare the performance of the prediction models, a total of six fusion models are designed for comparison experiments, namely TCN-BiGRU, CNN-SE-BiLSTM, CNN-SE-GRU, CNN-SE-LSTM, IWOA- BiLSTM and SSA-LSTM-GRU, where SE is squeeze-and-excitation networks \cite{Hu2020TPAMI}, which can significantly improve the interdependence type between the convolutional feature channels of the CNN module to enhance the representation capability of the network. Similar to the setup in subsection \ref{subsec:model_ablation}, a total of 12 sets of prediction results are obtained in this subsection by choosing 5\%, 15\%, and 25\% data scales as the prediction starting points. Taking the prediction results of battery Cell1 as an example, the model comparison prediction results for its three prediction starting points are shown in Fig. \ref{figur:diff_models}, from which it can be seen that comparing with other models, the SSA-BiGRU model achieves the best prediction results within their respective groupings.

Among all the compared models, the TCN-BiGRU model has the worst prediction results, failing to capture the decreasing trend of the SOH of the batteries. The reason is that for the TCN model, the time-convolution part of the model is more adept at dealing with cyclical data than the unidirectional trend of LiBs. For the three CNN fusion models with the addition of the SE module, the predictions between the prediction start point and the 40th cycle are poor, almost in a horizontal straight line, and the decreasing trend of the original data can be reflected gradually after 40 sets of cycles. In terms of side-by-side comparison, the SSA-BiGRU model gets a better fit in the pre-cycle and post-cycle of the test set, respectively, in the experiments with 5\% and 15\% prediction starting points.
Specifically, as shown in Figs. \ref{figur:f15a} and \ref{figur:f15b}, in the case of 5\% prediction starting point, it accurately predicts the decreasing trend of SOH before the 20th cycle, and gradually deviates from the true value after the 20th cycle; in the case of 15\% prediction starting point, it exhibits fluctuating nature in the early stage of prediction, and finally achieves accurate prediction in the end part of the cycle. When the proportion of training data reaches 25\% (i.e., the case of 25\% prediction starting point), as shown in FIg. \ref{figur:f15c}, all the models except the IWOA-BiLSTM model achieve effective improvement in prediction over the previous two sets of experiments, while the SSA-BiGRU model achieves the best fitting effect and achieves almost the same prediction as the original degradation trend, which is in line with the conclusion drawn in the subsection \ref{subsec:model_ablation}: when the proportion of training data is ratio increases to 25\%, the predictive ability of the SSA-BiGRU model is significantly improved.

\begin{figure}[!htbp]
	\centering
	\subfigure[Predictions starting point at 5\% (Cell3).]{\label{figur:f16a}	\includegraphics[width=0.32\textwidth]{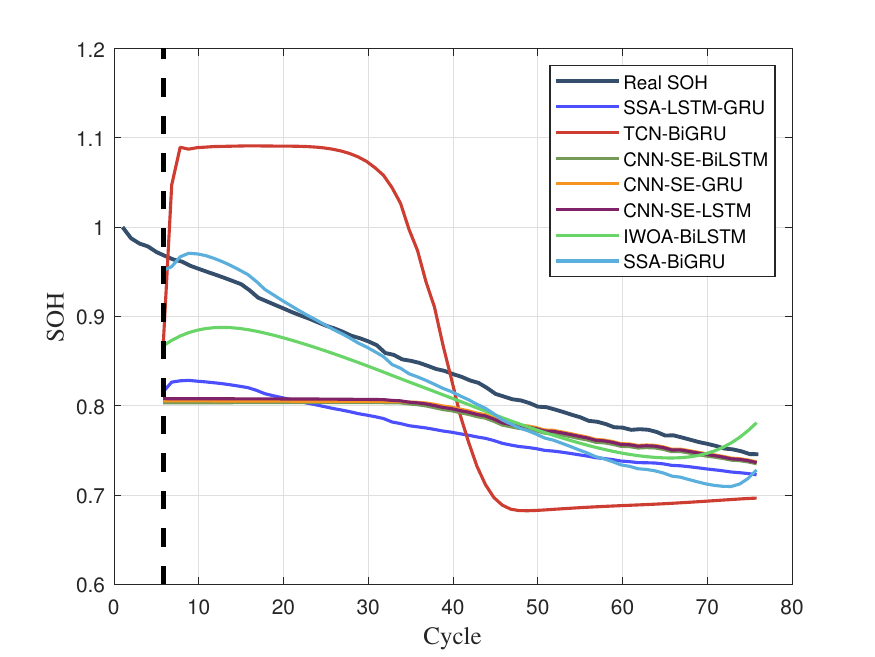}}
	\subfigure[Predictions starting point at 15\% (Cell3).]{\label{figur:f16b}	\includegraphics[width=0.32\textwidth]{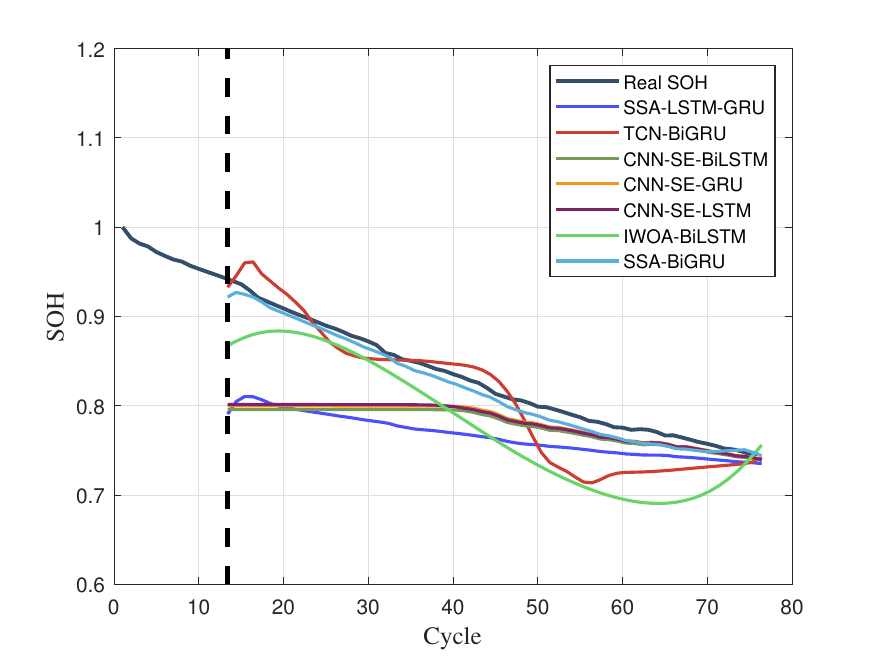}}
	\subfigure[Predictions starting point at 25\% (Cell3).]{\label{figur:f16c}	\includegraphics[width=0.32\textwidth]{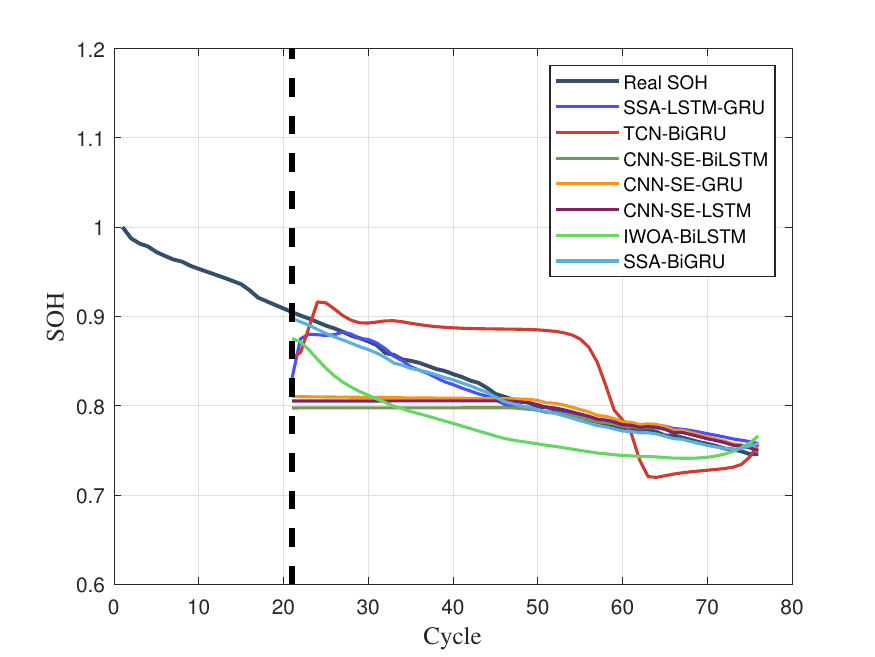}}
	\subfigure[Predictions starting point at 5\% (Cell7).]{\label{figur:f16d}	\includegraphics[width=0.32\textwidth]{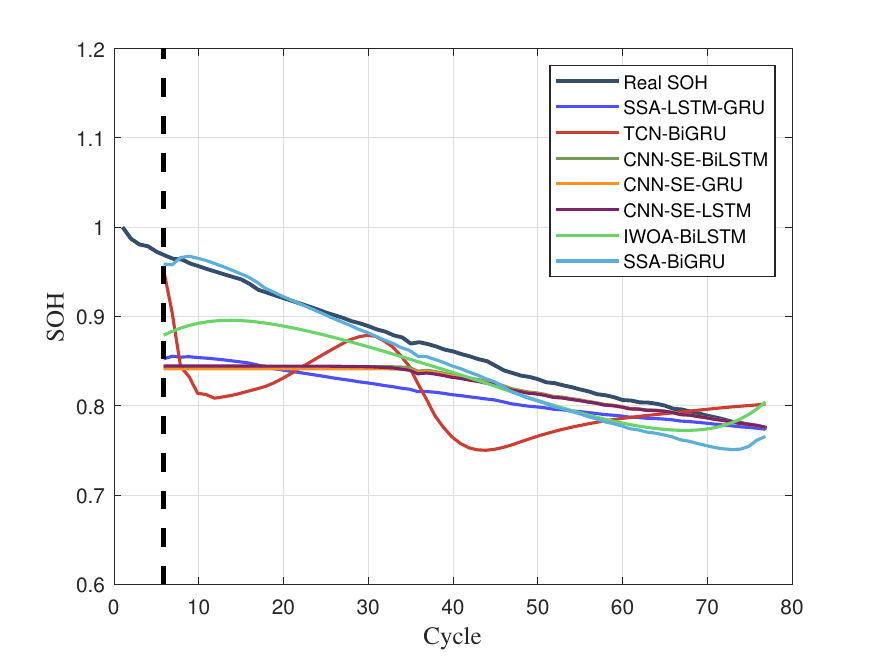}}
	\subfigure[Predictions starting point at 15\% (Cell7).]{\label{figur:f16e}	\includegraphics[width=0.32\textwidth]{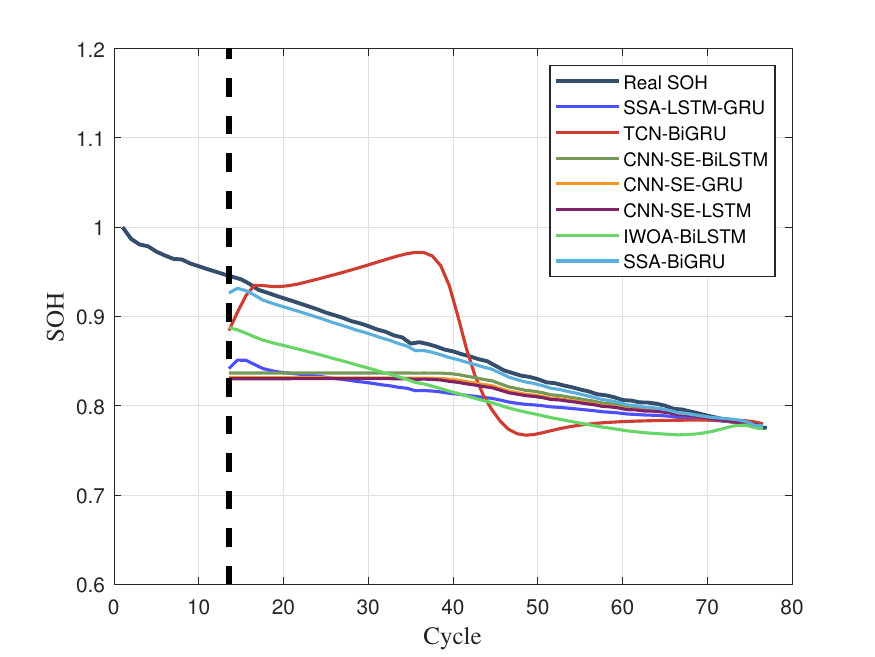}}
	\subfigure[Predictions starting point at 25\% (Cell7).]{\label{figur:f16f}	\includegraphics[width=0.32\textwidth]{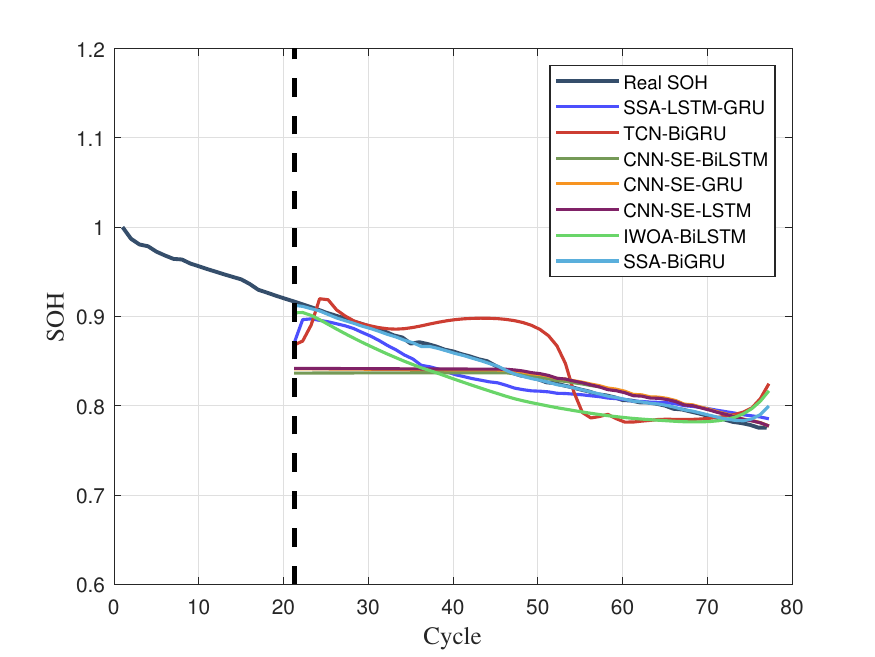}}
	\subfigure[Predictions starting point at 5\% (Cell8).]{\label{figur:f16g}	\includegraphics[width=0.32\textwidth]{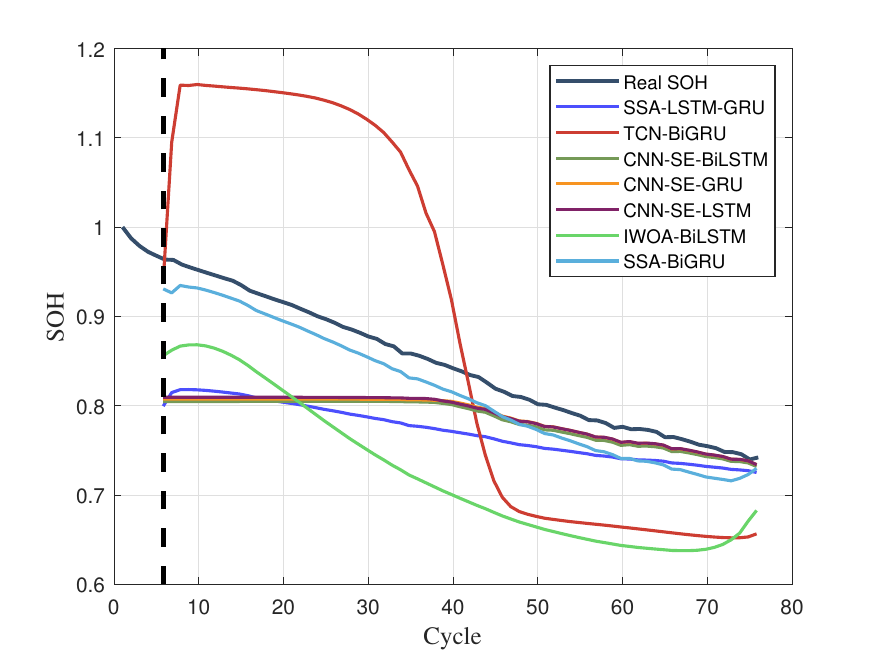}}
		\subfigure[Predictions starting point at 15\% (Cell8).]{\label{figur:f16h}	\includegraphics[width=0.32\textwidth]{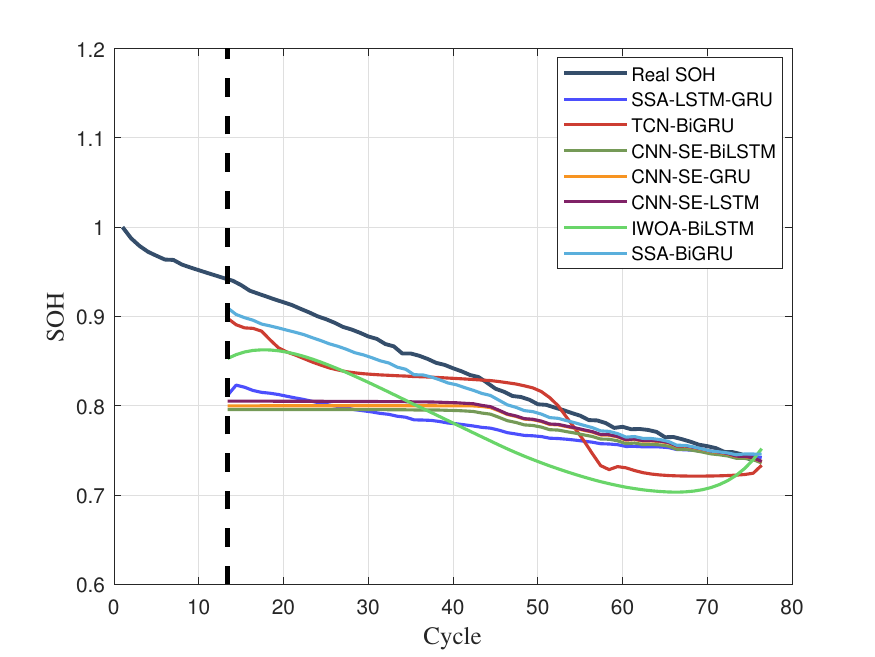}}
	\subfigure[Predictions starting point at 25\% (Cell8).]{\label{figur:f16i}	\includegraphics[width=0.32\textwidth]{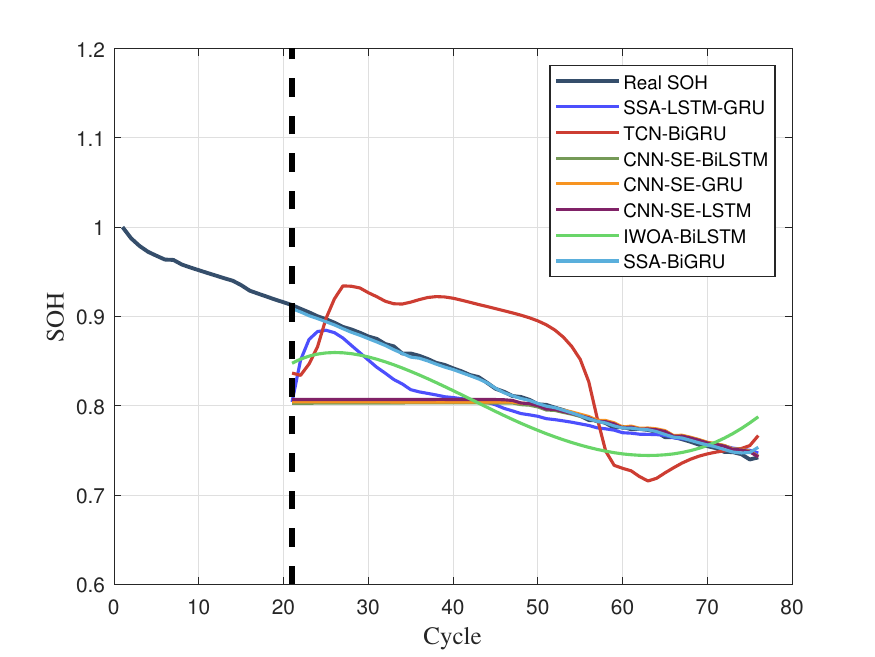}}
	\caption{SOH prediction results of different models for batteries Cell3, Cell7 and Cell8.}
	\label{figur:diff_models_diff_cells}
\end{figure}

To further validate the generalization of the proposed SSA-BiGRU model, the same comparison experiments on batteries Cell3, Cell7 and Cell8 are also performed in this subsection and the results are shown in Fig. \ref{figur:diff_models_diff_cells}. From Fig. \ref{figur:diff_models_diff_cells}, the following conclusions can be drawn: (1) The prediction results for each battery obey the conclusion presented earlier in this paper, i.e., the prediction results get better as the percentage of training data increases, and the prediction accuracy qualitatively changes when the percentage of training data reaches 25\% for the Oxford battery dataset; (2) In terms of the fit degree of the prediction values of the SSA-BiGRU model to the true values, Cell8 has a numerically poorer fit at the 5\% prediction starting point, but the overall trend still follows the degradation of the real SOH, and Cell7 at 15\% prediction starting point achieves the best fit result year-on-year; (3) Cell3 and Cell7 achieve a good fit at the beginning of prediction at the 5\% prediction start point, while the all three batteries achieve a good fit at the end of the prediction in the case of  15\% prediction start point and a high accuracy prediction at all stages in the case of 25\% prediction start point, indicating that the proposed SSA-BiGRU model has strong generalization when facing the battery data with similar degradation trends.

\begin{figure}[!htbp]
	\centering
	\subfigure[Prediction starting point at 5\% for Cell3.]{\label{figur:f17a}	\includegraphics[width=0.32\textwidth]{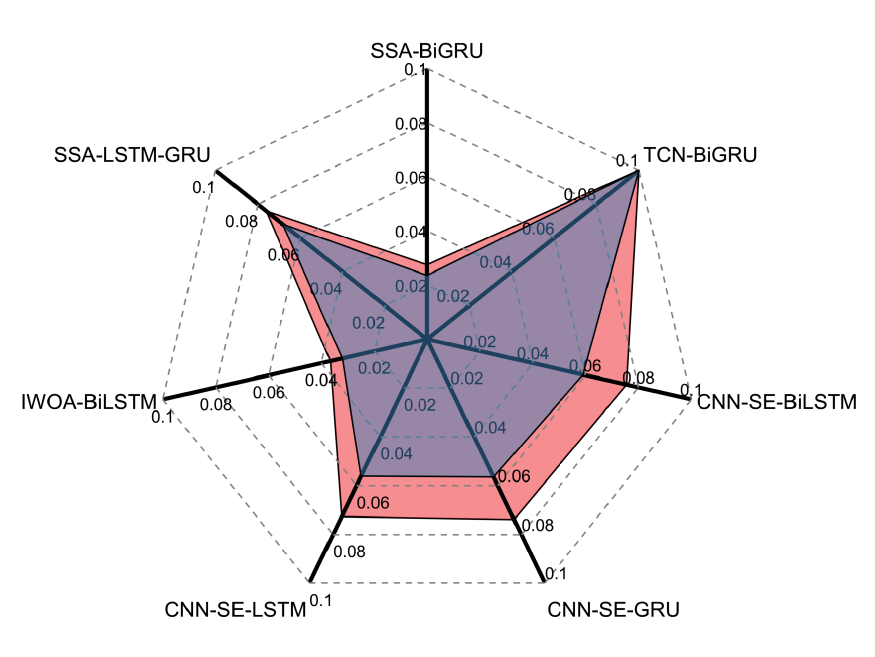}}
	\subfigure[Prediction starting point at 15\% for Cell3.]{\label{figur:f17b}	\includegraphics[width=0.32\textwidth]{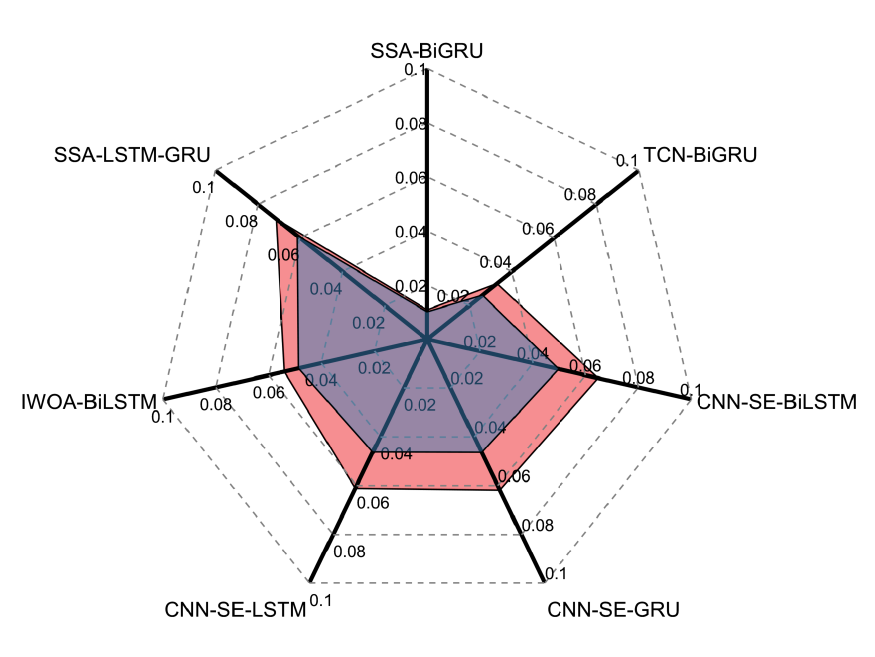}}
	\subfigure[Prediction starting point at 25\% for Cell3.]{\label{figur:f17c}	\includegraphics[width=0.32\textwidth]{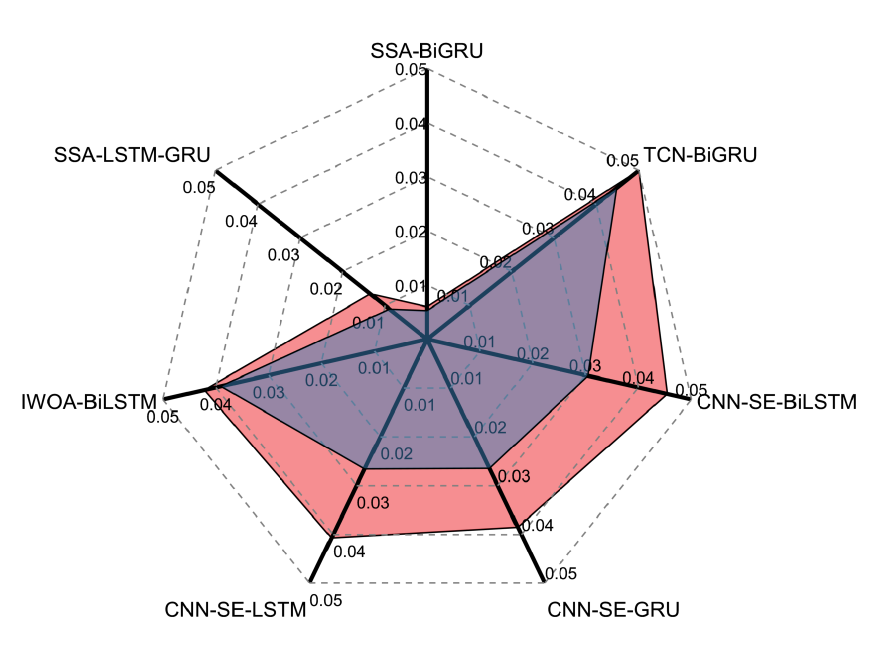}}
	\subfigure[Prediction starting point at 5\% for Cell7.]{\label{figur:f17d}	\includegraphics[width=0.32\textwidth]{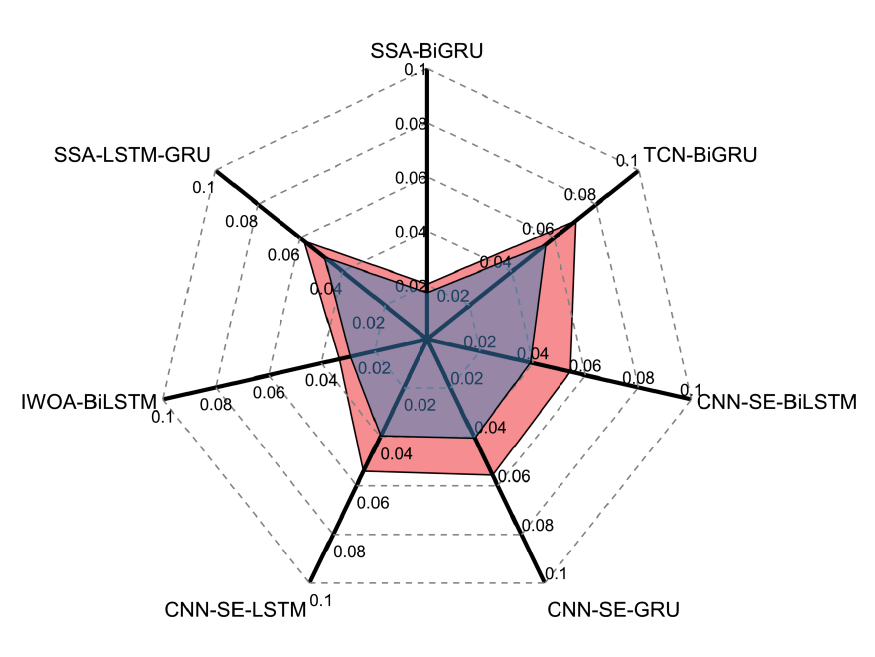}}
	\subfigure[Prediction starting point at 15\% for Cell7.]{\label{figur:f17e}	\includegraphics[width=0.32\textwidth]{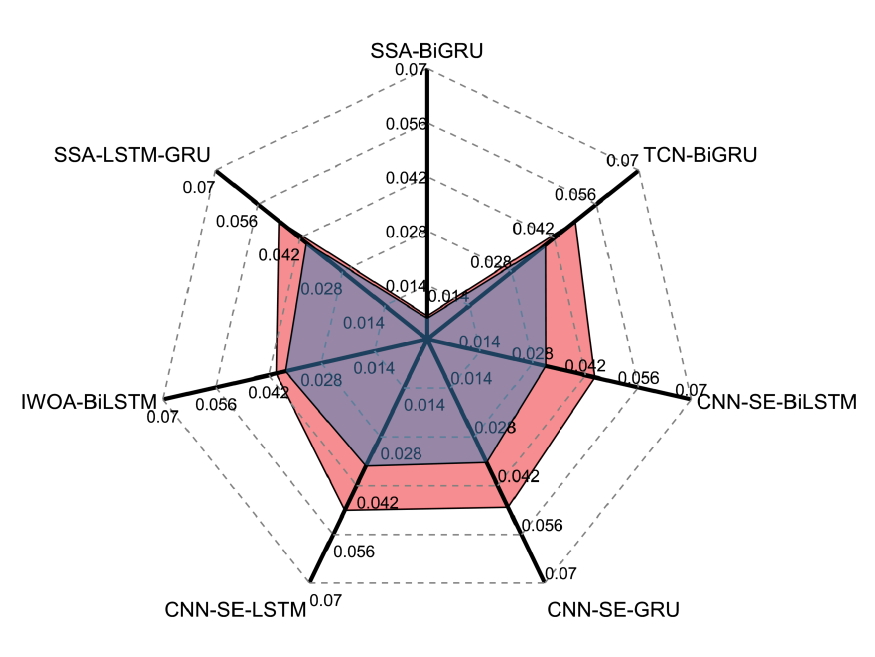}}
	\subfigure[Prediction starting point at 25\% for Cell7.]{\label{figur:f17f}	\includegraphics[width=0.32\textwidth]{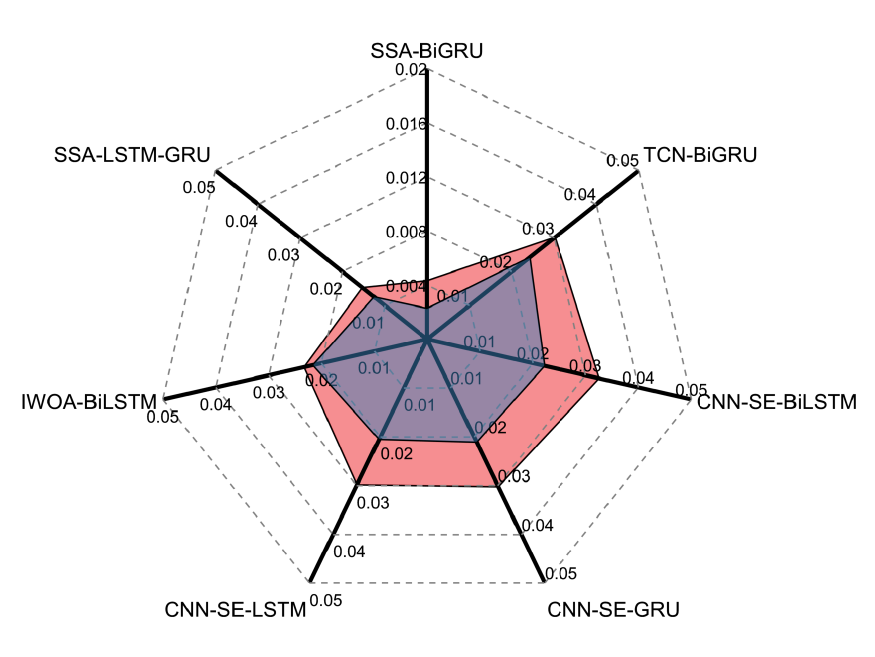}}
	\caption{Prediction errors for different prediction starting points for batteries Cell3 and Cell7.}
	\label{figur:diff_models_diff_batteries}
\end{figure}

Based on the evaluation indicators in subsection \ref{subsec:indicators}, the prediction errors using different models and different prediction starting points for the four groups of batteries are shown in Table \ref{Tab:t4}. In order to keep the table concise, the seven fusion models SSA-BiGRU, TCN-BiGRU, CNN-SE-BiLSTM, CNN-SE-GRU, CNN-SE-LSTM, IWOA-BiLSTM, and SSA-LSTM-GRU are sequentially denoted by FM1, FM2, FM3, FM4, FM5, FM6 and FM7. Moreover, the average values of different errors are calculated for different prediction starting points of a single battery. To improve the readability of the error data, the RMSE and MAE of Cell3 and Cell7 are visually represented in the form of radar charts, as shown in Fig. \ref{figur:diff_models_diff_batteries}, where red and blue represent RMSE and MAE, respectively.
\begin{table}[!htbp]
	\caption{Prediction errors for different models and different batteries.}
	\label{Tab:t4}
	\centering
	\tabcolsep=0.09cm
	{\footnotesize{
	\begin{tabular}{cccccccccccc}
		\toprule \multirow{2}{*}{Battery}&Prediction  &\multirow{2}{*}{Model}&\multirow{2}{*}{RMSE}&\multirow{2}{*}{MAE}&\multirow{2}{*}{MAPE}& \multirow{2}{*}{Battery}&Prediction &\multirow{2}{*}{Model}&\multirow{2}{*}{RMSE}&\multirow{2}{*}{MAE}&\multirow{2}{*}{MAPE}\\
		&starting point&&&&&&starting point&&&&\\
		\midrule
		\multirow{24}{*}{Cell1}&\multirow{8}{*}{5\%}&FM1&0.032169	&0.027687	&3.5169&\multirow{24}{*}{Cell3}&\multirow{8}{*}{5\%}&FM1&\textbf{0.027397}&	\textbf{0.023466}&	\textbf{2.9193}\\
		& &FM2&0.1664&0.15974	&19.5098&& &FM2&0.13025	&0.11977&	14.0291\\
			& &FM3&0.073522	&0.056576	&6.4548&& &FM3&0.075479	&0.05906&	6.65\\
			& &FM4&0.077462&	0.060072&	6.8606&& &FM4&0.073805	&0.056492	&6.3383\\
			& &FM5&0.068303	&0.051124	&5.8078&& &FM5&0.072628&	0.056173&	6.3134\\
			& &FM6&\textbf{0.029651	}&\textbf{0.022356}	&\textbf{2.6137}&& &FM6&0.036584&	0.032045&	3.7202\\
			& &FM7&0.084112	&0.074049	&8.6353&& &FM7&0.076053	&0.068161	&7.8494\\
			& &Avg&0.0759	&0.0645	&7.6284&& &Avg&0.0703&	0.0593&	6.8314\\
			\cline{2-6}
			\cline{8-12}
			&\multirow{8}{*}{15\%}&FM1&\textbf{0.015304	}&\textbf{0.012731}	&\textbf{1.5077}&&\multirow{8}{*}{15\%}&FM1&\textbf{0.010883}	&\textbf{0.010155}&	\textbf{1.2354}\\
			& &FM2&0.10146	&0.085608	&10.3431&& &FM2&0.032982&	0.0261	&3.2425\\
			& &FM3&0.05766	&0.04406	&5.1576&& &FM3&0.064785	&0.049747&	5.7075\\
			& &FM4&0.059359	&0.044316&	5.1716&& &FM4&0.061823	&0.046353	&5.3019\\
			& &FM5&0.055217&	0.040268	&4.6858&& &FM5&0.061117	&0.046226	&5.2936\\
			& &FM6&0.033392	&0.028809&	3.4587&& &FM6&0.054019	&0.048626&	5.9851\\
			& &FM7&0.062591	&0.052108	&6.1634&& &FM7&0.071265	&0.061214	&7.1306\\
			& &Avg&0.0550	&0.0440	&5.2126&& &Avg&0.0510&	0.0412	&4.8424\\
				\cline{2-6}
			\cline{8-12}
				&\multirow{8}{*}{25\%}&FM1&\textbf{0.0030443}&	\textbf{0.0020462}&	\textbf{0.26014}&&\multirow{8}{*}{25\%}&FM1&\textbf{0.0059948}&	\textbf{0.0053017}&	\textbf{0.63998}\\
			& &FM2&0.032167&	0.028426	&3.5776&& &FM2&0.051226	&0.044713	&5.5258\\
			& &FM3&0.041323	&0.028189	&3.3498&& &FM3&0.045424	&0.030463	&3.5372\\
			& &FM4&0.040309	&0.027252	&3.2378&& &FM4&0.038487	&0.026486	&3.0915\\
			& &FM5&0.037372	&0.024997	&2.9697&& &FM5&0.04067	&0.026604	&3.0853\\
			& &FM6&0.03096	&0.028454	&3.5144&& &FM6&0.042168	&0.03904	&4.73\\
			& &FM7&0.018568	&0.014428	&1.7534&& &FM7&0.013425	&0.0089509	&1.0901\\
			& &Avg&0.0291	&0.0220	&2.6661&& &Avg&0.0339	&0.0259	&3.09998\\
			\midrule
			\multirow{24}{*}{Cell7}&\multirow{8}{*}{5\%}&FM1&\textbf{0.020249}&	\textbf{0.017041}&	\textbf{2.061}&\multirow{24}{*}{Cell8}&\multirow{8}{*}{5\%}&FM1&\textbf{0.027074}&	\textbf{0.026429}&	\textbf{3.1917}\\
			& &FM2&0.070241	&0.05598	&6.3152&& &FM2&0.16783	&0.15446	&18.0154\\
			& &FM3&0.054168	&0.039037	&4.2788&& &FM3&0.076122	&0.059397	&6.6561\\
			& &FM4&0.055741	&0.040475	&4.4399&& &FM4&0.073982	&0.056423	&6.299\\
			& &FM5&0.054125	&0.039559	&4.3426&& &FM5&0.072794	&0.055613	&6.2103\\
			& &FM6&0.033247	&0.029053	&3.3008&& &FM6&0.12009	&0.11808	&14.1831\\
			& &FM7&0.058101	&0.048359	&5.3942&& &FM7&0.081822	&0.071773	&8.1976\\
			& &Avg&0.04941	&0.038501	&4.3046&& &Avg&0.088531	&0.077454	&8.9647\\
			\cline{2-6}
			\cline{8-12}
			&\multirow{8}{*}{15\%}&FM1&\textbf{0.0060757}&	\textbf{0.0054396}&	\textbf{0.628}&&\multirow{8}{*}{15\%}&FM1&\textbf{0.019754}&	\textbf{0.017125}&	\textbf{1.9884}\\
			& &FM2&0.04892	&0.039088	&4.549&& &FM2&0.036854	&0.032626	&3.8973\\
			& &FM3&0.044474	&0.031417	&3.5118&& &FM3&0.068354	&0.052809	&6.026\\
			& &FM4&0.048202	&0.035362	&3.9667&& &FM4&0.064915	&0.048402	&5.4964\\
			& &FM5&0.049081	&0.036377	&4.0846&& &FM5&0.061791	&0.045967	&5.2192\\
			& &FM6&0.039997	&0.037524	&4.3526&& &FM6&0.059819	&0.057968	&6.9727\\
			& &FM7&0.048972	&0.039872	&4.521&& &FM7&0.065925	&0.054347	&6.2535\\
			& &Avg&0.040817	&0.032154	&3.6591&& &Avg&0.053916	&0.044178	&5.1219\\
			\cline{2-6}
			\cline{8-12}
			&\multirow{8}{*}{25\%}&FM1&\textbf{0.0043176}&	\textbf{0.0022819}&	\textbf{0.27501}&&\multirow{8}{*}{25\%}&FM1&\textbf{0.0026961}&	\textbf{0.0019095}&	\textbf{0.23334}\\
			& &FM2&0.030188	&0.0242	&2.8933&& &FM2&0.059316	&0.052213&	6.3387\\
			& &FM3&0.032585	&0.022214	&2.5316&& &FM3&0.046406	&0.030678	&3.5287\\
			& &FM4&0.030233	&0.021064	&2.4065&& &FM4&0.045627	&0.030088	&3.4612\\
			& &FM5&0.029821	&0.020492	&2.3377&& &FM5&0.043896	&0.02853	&3.2768\\
			& &FM6&0.023261	&0.021614	&2.576&& &FM6&0.029532	&0.027418	&3.3404\\
			& &FM7&0.015276	&0.012525	&1.4683&& &FM7&0.025794	&0.01861	&2.1922\\
			& &Avg&0.023669	&0.017770	&2.0698&& &Avg&0.036181	&0.02706	&3.1959\\
		\bottomrule
	\end{tabular}}}
\end{table}

\textbf{Qualitative analysis}. From Table \ref{Tab:t4}, it can be seen that the FM1 model obtains the best prediction result and is below the average in all cases except for battery Cell1 at 5\% prediction starting point. In the experiments for battery Cell1, the FM1 model only performs weaker than the FM6 model in the experiments of 5\% prediction starting point, and still achieves the optimal prediction in the experiments of 15\% and 25\% training data, which corroborates the previous conclusions of this paper that individual models may perform well in specific cases but lack generalization. For individual battery experiments, the proposed SSA-BiGRU model achieves optimal prediction results at 25\% prediction starting point, demonstrating that the HI extracted from the IC curves can well characterize the degradation trend of LiBs, allowing the model to predict the actual SOH of LiBs after ‘learning’ the information, and confirming the previous conclusions that the HI extracted from the IC curves can be used to predict the actual SOH of LiBs. This also confirms the generality of the previous conclusions to the Oxford battery dataset. In addition, the later the prediction starting point is set, the more degradation trend information can be learn, and thus the prediction model can learn more information to predict the SOH in the later stage more accurately, which is more important than the early stage prediction for battery health management, as the later SOH prediction can provide guidance for the timing of battery replacement.

\textbf{Quantitative analysis}. Taking the RMSE of four groups of batteries as an example, the following results can be obtained: (1) 5\% prediction starting point: Cell8 reduces the most compared to the mean value, about 69.42\%, while Cell1 reduces the least, 57.62\%; (2) 15\% prediction starting point: Cell7 reduces the most compared to the mean value, about 85.11\%, while Cell8 reduces the least, 63.36\%; (3) 25\% prediction starting point: Cell8 reduces the most compared to the mean value, about 92.55\%, while Cell7 reduces the least, 81.76\%. For the MAE of the SSA-BiGRU model in the four sets of batteries, the prediction accuracy improvement of Cell1, Cell3, Cell7 and Cell8 for 5\% to 15\% and 15\% to 25\% of the prediction starting point are, in order, 54.02\% vs. 83.93\%, 56.72\% vs. 47.79\%, 68.08\% vs. 58.05\% and 35.2\% vs. 88.85\%. It can be seen that the improvement rate of prediction accuracy of Cell1 and Cell8 increases with the increase of prediction starting point, which is in line with the logical cognition; the decrease of prediction accuracy improvement rate of Cell3 and Cell7 is because they have already achieved higher prediction accuracy with 15\% prediction starting point. 

From the qualitative analysis and quantitative analysis, the SSA-BiGRU model proposed in this paper is the best prediction model for predicting the SOH of LiBs under driving conditions.

\subsection{SOH prediction of real EV road driving conditions based on SSA-BiGRU model}
Although the Oxford battery dataset was collected under simulated urban driving conditions, it still lacks sufficient realism of driving conditions. Therefore, in this section, the SSA-BiGRU model is applied to the dataset of real EVs with on-road driving conditions shown in subsection \ref{subsubsec:real_ev} for SOH prediction to validate the real-world prediction capability and generalization of the SSA-BiGRU model.

\begin{figure}[!htbp]
	\centering
	\includegraphics[width=1\textwidth]{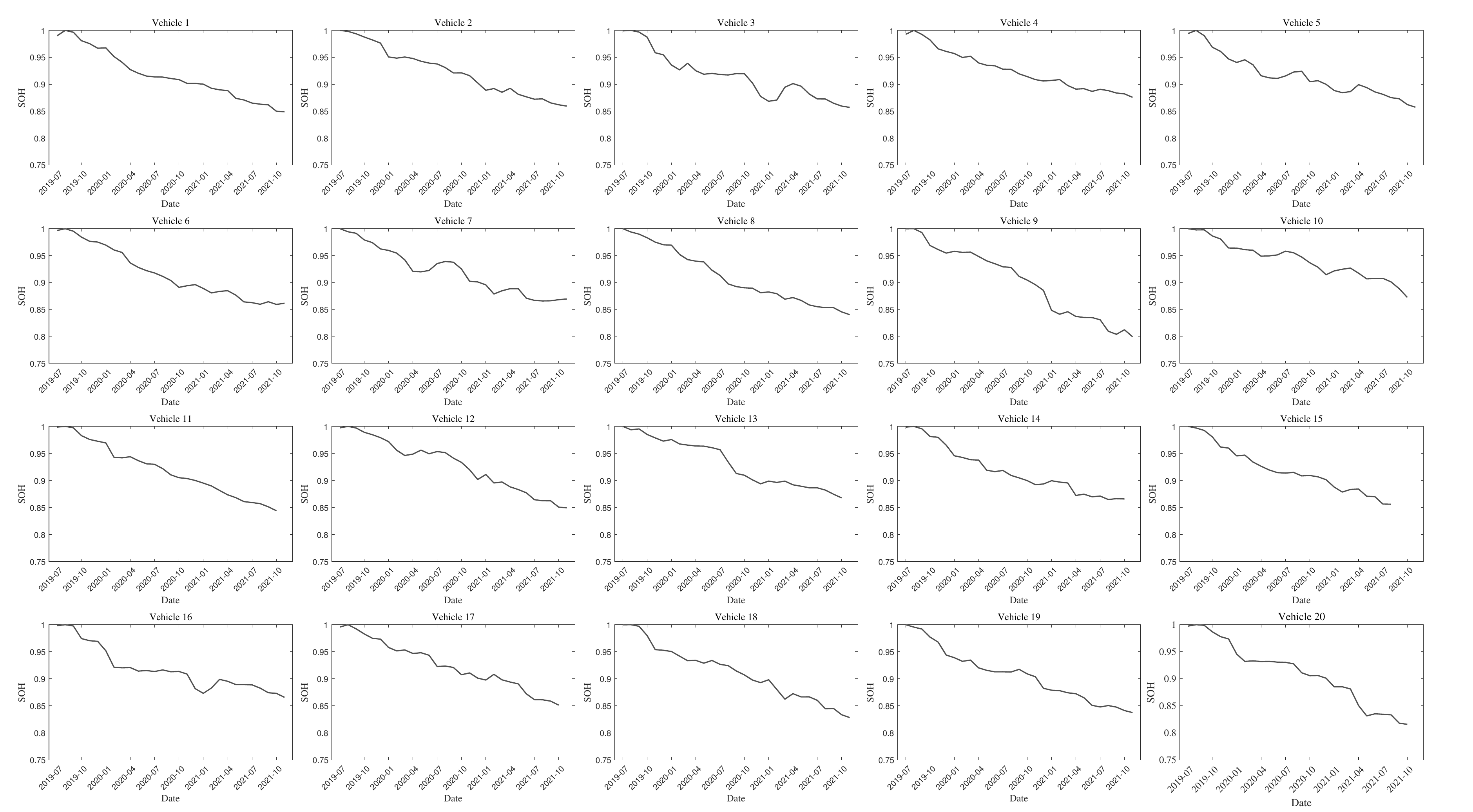}
	\caption{SOH of 20 EVs charging data by using median.}
	\label{figur:f18}
\end{figure}

The median of 20 EVs charging data is selected as the test data in this subsection, and their SOHs are shown in Fig. \ref{figur:f18}. It is worth noting that the degradation recorded in this dataset is calendar aging-based. Since the median of each month's charging data of the EVs is chosen as the charging capacity of the month, it must be numerically smaller than the rated capacity of the battery, i.e., 145 Ah. Moreover, since there are multiple incomplete charging and discharging behaviors under actual driving conditions, if 145 Ah is used as the denominator for SOH calculation, the starting SOH of the battery is not 1, which is not reasonable in practice. Therefore, we choose to use the maximum value of the median of each vehicle itself as the denominator of SOH in this section.

\begin{figure}[!htbp]
	\centering
	\includegraphics[width=1\textwidth]{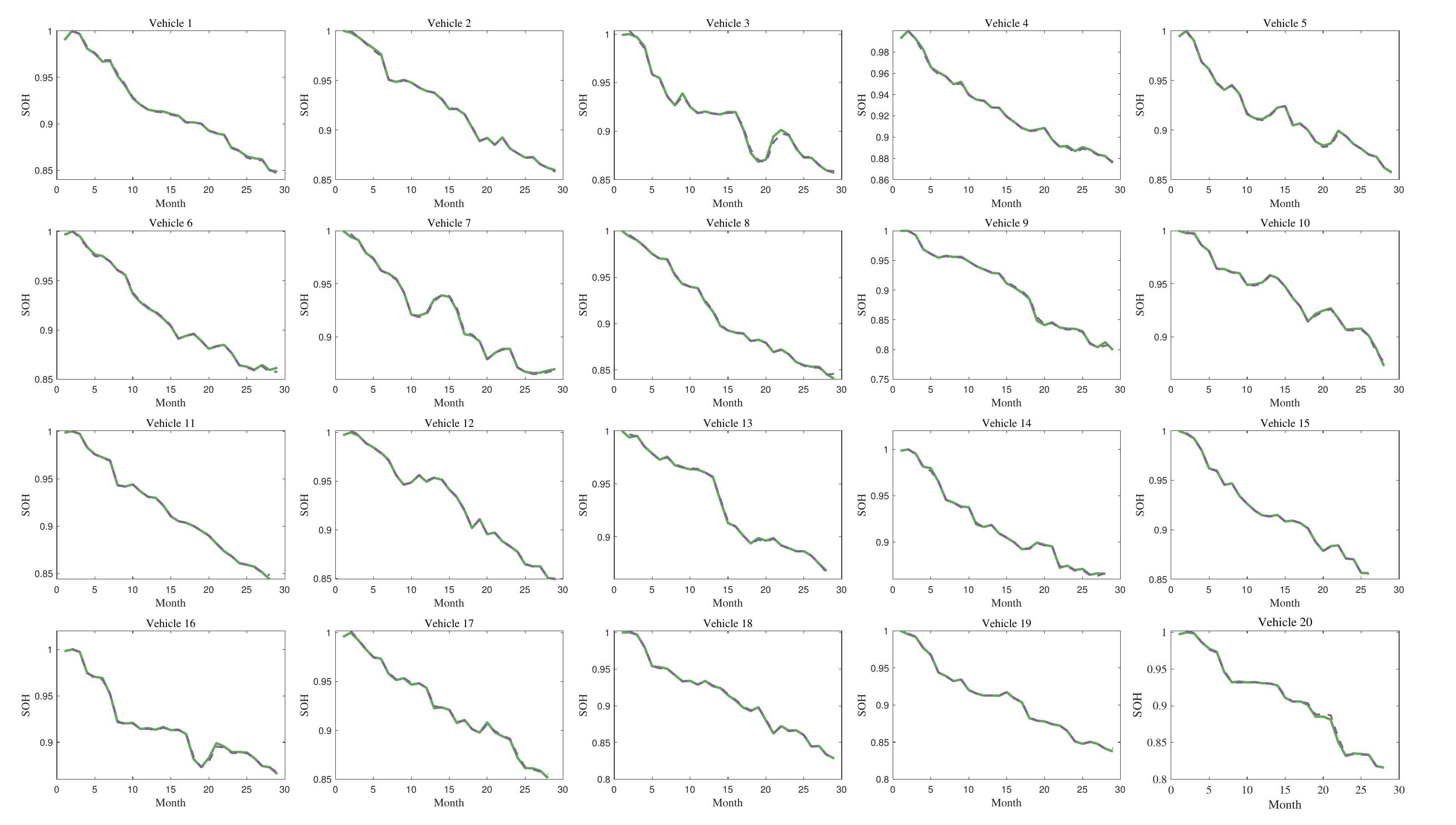}
	\caption{SOH prediction of 20 EVs when the prediction starting point is the second month.}
	\label{figur:f19}
\end{figure}

In the experiment, in order to more closely match the practical application scenarios, the following situation is defined: the complete SOH degradation data of one EV in the same batch of EVs of the same model is known, which is used for the SOH prediction of the remaining EVs after training the prediction model. Specifically, the SOH of V1 is chosen as the known SOH to be used as the training set of the SSA-BiGRU model, and the SOHs of V2 to V20 are used as the test set, respectively, with the prediction starting point in the second month. In addition, to ensure the completeness of the experiment, the data of an EV is randomly selected from the data other than V1 for model training to predict the SOH of V1, and the final randomly selected data of an EV is V17. The experiment results of prediction starting point is 2 are shown in Fig. \ref{figur:f19}, where the green solid line is the real SOH, while the purple dashed line is the prediction value. In order to facilitate the description, the horizontal axis is changed from a specific point in time to the month.

As can be seen from Fig. \ref{figur:f19}, when the training set is the complete SOH degradation data, the SSA-BiGRU model has good prediction results on data with the same decay trend, and the prediction values can fit the true values well. Although the prediction values in the last month of individual EV will be slightly higher than the true values, it is still within the controllable range. Qualitatively, the prediction results show that the SSA-BiGRU model has high prediction accuracy in SOH prediction for real working conditions and has good generalization with a fixed training set. The errors for the prediction starting point of the second month are shown in Table \ref{Tab:t5}.
\begin{table}[!htbp]
	\caption{Prediction errors of 20 EVs for the prediction starting point of the second month.}
	\label{Tab:t5}
	\centering
	\begin{tabular}{cccc}
		\toprule
		Vehicle & RMSE       & MAE        & MAPE     \\
		\midrule
		V1          & 0.0011081  & 0.00092865 & 0.10203  \\
		V2          & 0.00076867 & 0.00057855 & 0.061831 \\
		V3          & 0.0021389  & 0.0015506  & 0.16988  \\
		V4          & 0.00094467 & 0.00074527 & 0.080543 \\
		V5          & 0.00097924 & 0.00078056 & 0.085334 \\
		V6          & 0.0011892  & 0.00091188 & 0.10009  \\
		V7          & 0.0011776  & 0.00092561 & 0.10045  \\
		V8          & 0.0013306  & 0.00086904 & 0.097033 \\
		V9          & 0.0028611  & 0.0017309  & 0.20326  \\
		V10         & 0.0012015  & 0.00088157 & 0.095373 \\
		V11         & 0.0008931  & 0.00042138 & 0.047149 \\
		V12         & 0.00066851 & 0.00058198 & 0.06235  \\
		V13         & 0.001037   & 0.00075801 & 0.081223 \\
		V14         & 0.0010537  & 0.00074568 & 0.081889 \\
		V15         & 0.00037008 & 0.00026432 & 0.028793 \\
		V16         & 0.0014532  & 0.0010986  & 0.12058  \\
		V17         & 0.0013576  & 0.0011037  & 0.12147  \\
		V18         & 0.0015775  & 0.00096838 & 0.10872  \\
		V19         & 0.00092765 & 0.00048938 & 0.055308 \\
		V20         & 0.0024002  & 0.0016193  & 0.18365 \\
		\bottomrule
	\end{tabular}
\end{table}

From Table \ref{Tab:t5}, the prediction errors of SOH for EVs based on the SSA-BiGRU model can all be kept at a low level. The RMSE, for example, has a mean value of 0.0013, which is a stronger prediction than the performance of this model on the Oxford battery dataset in the subsection \ref{subsec:oxford_prediction}. The model exhibits relatively large errors on V3, V9 and V20, but remains below 0.003. As a whole, the best prediction accuracy is achieved by V15, which demonstrates the lowest RMSE, MAE, and MAPE. Moreover, although V2, V11, V12, and V19 also have relatively low error, suggesting that their predictions are more accurate, but there are still gaps compared to V15.

\begin{figure}[!htbp]
	\centering
	\includegraphics[width=1\textwidth]{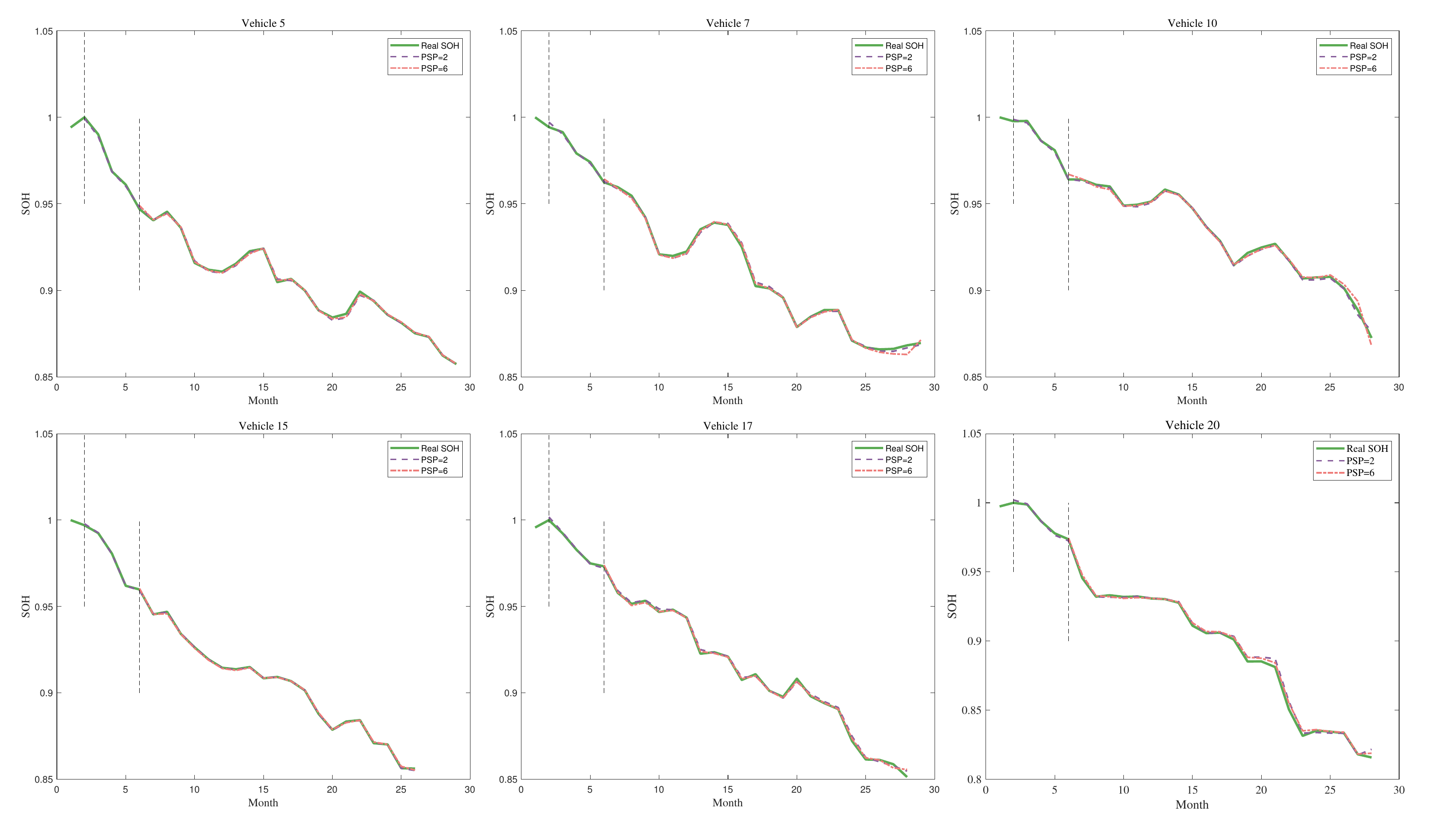}
	\caption{Prediction results of different prediction starting points for 6 EVs.}
	\label{figur:f20}
\end{figure}

In order to verify the prediction ability of the SSA-BiGRU model for different prediction starting points (months), a total of six sets of battery data, V5, V7, V10, V15, V17 and V20, are selected for the experiments. Considering that the degradation data belongs to the calendar aging mode, which is short in length itself, the prediction starting point is set to be the 6th month in order to avoid the prediction starting point being backward leading to the actual SOH exceeding the failure threshold. To further verify the generalization of the SSA-BiGRU model, the degradation data of V1 is still chosen for the training set. To facilitate the comparison between the errors, some of the errors contained in Table \ref{Tab:t5} are multiplexed. As a result, the SOH prediction errors at different prediction starting points for the six sets of batteries are shown in Table \ref{Tab:t6}, and the prediction results are shown in Fig. \ref{figur:f20}, where the PSP (prediction starting point) in the legend stands for the prediction starting point, which shows that the prediction results are more accurate, but there is still a gap compared with V15.

\begin{table}[!htbp]
	\caption{ Prediction errors for different prediction starting points for 6 EVs.}
	\label{Tab:t6}
	\centering
	\begin{tabular}{ccccc}
		\toprule
		Vehicle        & Prediction starting point (Month) & RMSE       & MAE        & MAPE     \\
			\midrule
		\multirow{2}{*}{V5}  & 2                                 & 0.00097924 & 0.00078056 & 0.085334 \\
		& 6                                 & 0.00084894 & 0.00064002 & 0.070263 \\
		\multirow{2}{*}{V7}  & 2                                 & 0.0011776  & 0.00092561 & 0.10045  \\
		& 6                                 & 0.0015771  & 0.0011119  & 0.12374  \\
		\multirow{2}{*}{V10} & 2                                 & 0.0012015  & 0.00088157 & 0.095373 \\
		& 6                                 & 0.0017746  & 0.0011903  & 0.13001  \\
		\multirow{2}{*}{V15} & 2                                 & 0.00037008 & 0.00026432 & 0.028793 \\
		& 6                                 & 0.00053876 & 0.00042765 & 0.047537 \\
		\multirow{2}{*}{V17} & 2                                 & 0.0013576  & 0.0011037  & 0.12147  \\
		& 6                                 & 0.0012577  & 0.00086552 & 0.09677  \\
		\multirow{2}{*}{V20} & 2                                 & 0.0024002  & 0.0016193  & 0.18365  \\
		& 6                                 & 0.0019494  & 0.0015021  & 0.17076 \\
		\bottomrule
	\end{tabular}
\end{table}

From Table \ref{Tab:t6} and Fig. \ref{figur:f20}, as a whole, the prediction errors for the three groups of EVs with a prediction start point of the 6th month are lower than the 2nd month, while the other three groups of EVs, V7, V10 and V15, show an increasing trend. For the different EVs, V15 outperforms the others on all evaluation indicators, indicating higher predictability and stability. On the other hand, V20  remains consistent with the results presented in Table \ref{Tab:t5}, with a relatively high level of error. Additionally, Table \ref{Tab:t6} shows that when the proposed SSA-BiGRU model is applied to the degradation prediction of real driving data, it does not follow the conclusion of the non-real driving data, i.e., the prediction accuracy of the SSA-BiGRU model rises as the prediction starting point is shifted back. This is because in the experiments of this subsection, the training data was always fixed, i.e., the degradation data of V1, and the factors that changed were limited to the prediction start points for different EVs. When the test data is input into the prediction model, the SSA-BiGRU model will predict the SOH of the corresponding EV from the degradation trend learnt from V1. Therefore, the SSA-BiGRU model proposed in this paper is more suitable for early degradation prediction of batteries of other EVs in real working conditions when the complete battery degradation data of the EV is inputted, which facilitates the timely replacement of batteries and reduces the failure rate and the degree of danger.
\section{Conclusions and further works}\label{Sec:Cons}
In this paper, we proposed a strategy for SOH prediction of LiBs under driving conditions based on the extraction of one-dimensional HI from the IC curve, combining full parameter domain SSA with a dual-module BiGRU. The proposed SSA-BiGRU model makes full use of the advantages of bidirectional learning sequence features of BiGRU, and the SSA is used to perform full-dimensional parameter search for the dual-module BiGRU to keep the prediction model in the optimal prediction state. The proposed SSA-BiGRU model is tested based on the Oxford battery dataset, which shows its excellent prediction ability and generalization on the dataset of simulated driving conditions. On this basis, the SSA-BiGRU model is tested on the charging dataset of 20 EVs with real driving conditions, the median monthly charging capacity is selected as the experiment subject, and the data of a fixed single EV is used as the training set with a ratio of 1 to 19. The robustness of the SSA-BiGRU model under real driving conditions is verified, and is proved to have high prediction accuracy. In addition, the experiments show that the SSA-BiGRU model is more suitable for the early SOH prediction of EV power battery in real working conditions, and its lowest RMSE can reach 0.00037008 and the highest is not more than 0.0025 when the prediction starting point is 2, which can provide early and high accuracy SOH prediction.

In this paper, the SSA-BiGRU model is used to predict the SOH of LiBs for driving conditions, and future work is needed to improve the prediction accuracy of this model with a small amount of training data as well as its early prediction ability. In SOH prediction based on real EV charging data, the complete degradation data of an EV needs to be predicted in advance, which requires high data acquisition in real engineering. In future work, the establishment of a large model for SOH prediction of LiBs based on different EVs and different operating conditions will be considered, and multiple datasets will be used to divide the Token to train the prediction model, which will be further deployed on battery data from different devices, batches, operating conditions, and usage habits to achieve real-time SOH prediction by taking advantage of cloud computing \cite{drake2014cloud} and the internet of things \cite{perkel2017internet}. 

\section*{Declaration of competing interest}
 The authors declares that they have no known competing financial interests or personal relationships that could have appeared to influence the work reported in this paper.
 
%

\printcredits

\bibliographystyle{unsrt}

\bibliography{RESS}


%
%
%

\end{document}